\documentclass{aa}
\usepackage{epsfig}
\usepackage{graphicx}
\begin{document}

\title{Evidence for a deficit of young and old stars in the Milky Way inner in-plane disc}

\author{M. L\'opez-Corredoira\inst{1}, A. Cabrera-Lavers\inst{2}, O. E.
Gerhard\inst{1}, F. Garz\'on\inst{2,3}}
\institute{$^1$ Astronomisches Institut der Universit\"at Basel,
Venusstrasse 7, CH-4102 Binningen, Switzerland\\
$^2$ Instituto de Astrof\'\i sica de Canarias, C/ V\'\i a L\'actea, s/n,
E-38200 La Laguna, Tenerife, Spain\\
$^3$ Departamento de Astrof\'{\i}sica, Universidad de La Laguna, E-38200 La Laguna, Tenerife, Spain}

\offprints{acabrera@ll.iac.es}

\date{Received xxxx / Accepted xxxx}

\abstract{
We give  independent proof of the deficit
of stars in the in-plane central disc ($2<R<4$ kpc, 
$|b|\la 3^\circ $) with respect to the predictions of a pure exponential 
 density distribution.\\
We use three different methods: 1) the inversion of the red clump giant distribution
in near-infrared colour--magnitude diagrams  to obtain the star 
density along the line of sight; 2) the determination of the density distribution
of 1612 MHz sources by means of the distance 
determination of OH/IR sources from their 
kinematical information; 3) an analysis of near-
 and mid-infrared star counts and
comparison with models. All the tests give the same result: a deficit of
stars in the inner disc with respect to an exponential disc (either with constant
scaleheight or
extrapolated from the outer disc), but only in near plane regions ($|b|\la 3^\circ $).
This deficit  might be interpreted as a flare in the vertical distribution.
The in-plane density is almost independent of 
$R$ and not an exponential law of the type $\rho \propto \exp (-R/h)$.
Further away from the plane, however, the density 
 increases towards the centre due to the increase of the scaleheight.
 Tests also show that this result cannot be due to  extinction.
This deficit affects both  the young  and the old populations,
so it is probably a rather stable feature of the disc, and might be
due to the existence of an in-plane bar  sweeping the near-plane stars.\\
An approximate expression of the disc density within
$2<R<8$ kpc is:
$\rho(R,z)\propto e^{-\left(\frac{R}{1970\ {\rm pc}}+
\frac{3740\ {\rm pc}}{R}\right)}e^{\frac{-|z|}{h_z(R)}}$, with $h_z(R)\approx 285[1+0.21\ 
{\rm kpc}^{-1}\ (R-R_\odot)+0.056\ {\rm kpc}^{-2}\ (R-R_\odot)^2]
\ {\rm pc}$. 
}

\authorrunning{L\'opez-Corredoira et al.}

\titlerunning{deficit of stars in the Milky Way inner in-plane disc}

\maketitle

\section{Introduction}

L\'opez-Corredoira et al. (2002, hereafter L02)  obtained
a model of the outer parts of the stellar disc 
(galactocentric radius $R$ greater than 6 kpc), based on the 2MASS near
infrared point sources survey.
Among their findings are the stellar warp coincident with the gas warp, 
the flare, the absence of an external cut-off out to $R=15$ kpc and the measurement of 
the different parameters of the old population density distribution,
generally in agreement with other independent studies.
Here, we  continue the analysis of the disc structure in the 
 inner stellar disc, examining whether or not this corresponds  to an
extrapolation of the L02 disc model towards the centre and, if not, 
what  the differences are.

A major concern is whether  extrapolation of the exponential 
in the inner stellar disc is appropriate or whether there is some deficit
of stars with respect to this distribution.
It is well known that many barred galaxies have holes
(Ohta et al.\ 1990); up to half of them, according to Anderson et al.\ (2002)
based on  analysis of the bulge-disc decomposition carried out in Baggett et al.\ (1998). 
In a large number of external galaxies, due to their more favourable 
orientation, the central hole in the disc has been detected in the 
distribution of molecular and atomic gas. See, for example, 
Aguerri et al.\ (2001), who describe a hole 
in the HI distribution in NGC 5850 associated with the central bar region, 
as is the case presented by Bottema \& Verheijen (2002) in NGC 3992. 
These holes have been attributed to the presence of gas velocities 
perpendicular to the Galactic plane (Sancisi 1999). However, 
there are also cases where the CO distribution, used to 
infer the HII density, does not reveal such a hole, as is the case, for example, of Sakamoto 
et al.\ (1999), whose sample is somewhat biased towards the non-hole 
gas distribution because of  selection criteria.
The hole does not usually consist strictly of an abrupt
truncation of the disc. It can represent a gradual density decrease
(for instance, Freeman type II truncation; Freeman 1970), instead of increasing 
towards the center within a certain radius, or even a constant
density or increasing density but with a slope much lower than that produced by
the extrapolated exponential law.

In our Galaxy, this hole (from now on, we will call the hole a
``deficit'' of gas or stars since, strictly speaking, the hole is not
totally empty;  ``deficit'' here is always with respect to the extrapolation of
the purely exponential disc towards the inner radii)
is observed in the gas distribution
(i.e., Robinson et al.\ 1988). For the stellar distribution the answer
is not very clear at present. Several disc models have been produced using
COBE/DIRBE flux maps. Although the general features are approximately
coincident, it seems that there is no agreement regarding the
deficit of stars. For instance, Freudenreich (1998) and L\'epine \& Leroy (2000)
agree that there is a such a deficit, while
Bissantz \& Gerhard (2002, hereafter B02) 
can reproduce quite accurately the flux maps with an exponential disc
even in the inner parts.
Caldwell \& Ostriker (1981) gave 
a model of the disc that consists of a difference of two exponentials with different scalelengths, 
which gives a deficit of stars in the inner part (see fig. 15 in Caldwell
\& Ostriker 1981) based on kinematical measurements.
Microlensing experiments seemed to favour a deficit of stars (Kiraga \& 
Paczy\'nski 1994), but the poorly--known interstellar extinction
at optical wavelengths gives some uncertainties in their models.
Moreover, Kiraga et al. (1997) recognize that their previous
paper might contain some errors due to an incorrect identification of populations
in the optical colour--magnitude diagrams.
Analysis of OH/IR sources, both from the young and intermediate--old populations, led
Baud et al.\ (1981) to conclude that there must be a deficit of stars in the two
populations; these results are more convincing, and we think that
a corroboration of this result with newer surveys will 
be valuable (see \S \ref{.OHIR}).
This topic has so far not received much attention in terms of star counts. 
L\'opez-Corredoira et al.\ (2001, hereafter L01) have claimed that a deficit of
stars is necessary to reproduce some near-plane counts but 
they did not try to fit off-plane regions, and not too 
much attention was paid to the problem in that paper.
Precisely because of this, we wish to examine the possible deficit of stars in the inner disc, using point source catalogues, 
which contain more information than flux-maps alone.

A further area of study is the real nature of the deficit of stars, should this be genuine. 
 A flared stellar distribution places the stars at a higher elevation from the plane 
than that predicted from a pure exponential law, hence giving the impression of a hole in 
the plane. Flared structures are well recognized in the distribution of gas (see Burton \& 
Hekkert 1986) and stars (L02; Alard 2000, and references therein) in the outer disc. The 
presence of that type of morphology in the central disks of spiral galaxies is 
not so well studied due to the inherent difficulties in separating the several 
structural contributions to the observed counts or flux. A good example of that 
difficulty can be found in Nikolaev et al. (2004), who reported a symmetric 
warp in the inner disk, the central 4 degrees, of the LMC  based on accurate analysis 
of MACHO and 2MASS counts.

\section{Data}

The data for this work were taken from several sources: mainly 2MASS, MSX
and ATCA/VLA, and some other DENIS and TMGS data taken from a former 
paper (L01). The second release of the 2MASS project
(Skrutskie et al.\ 1997, http://www.ipac.caltech.edu/2mass/releases/docs.html)
provides us with some data in the  regions of interest to study
the inner disc isolated from other components ($1.5^\circ<|b|<6.5^\circ$,
$15^\circ<l<40^\circ$). We use these data to produce $K$-band star counts ($m_K$) 
and $m_K$ vs. $(J-K)$ colour--magnitude diagrams (CMDs) 
in the regions given in Table \ref{Tab:giants}. 
The MSX Point Source Catalogs (version 1.2; Egan 
et al.\ 1999) provides us with star counts in the  
$D$ (14.65 $\mu$m) mid-infrared band in the regions $|b|<0.5^\circ$, $|l|<45^\circ$.
OH/IR stars from the ATCA/VLA catalogue (Sevenster et al.\ 1997a,b, 2001)
in the radio maser line at 1612.23 MHz
at $|b|<2.5^\circ$, $|l|<45^\circ$ were used to obtain star counts
and the distribution of the sources as a function of galactocentric radius.

\begin{table}[!h]
\caption{Selected regions used to extract the colour--magnitude diagram
from 2MASS sources so as Ato apply the method of \S \ref{.CMD}}
{\centering \begin{tabular}{ccc}\hline 
\hline
{$l$ (deg)}&
{$b$ (deg)}&
{area (deg$^2$)}\\
\hline 
15 & 2 & 0.57  \\
15 & 3 & 1.00 \\
15 & $-$3 & 1.00 \\
15 & $-$6 & 1.00 \\
18 & $-$2 & 0.92  \\
18 & $-$3 & 0.97  \\
20 & $-$3 & 1.00 \\
20 & $-$6 & 1.00 \\ \hline
\label{Tab:giants}
\end{tabular}\par}
\end{table}

\section{The inner disc from CMD analysis}
\label{.CMD}

\subsection{Method}

\begin{figure}[!h]
{\par\centering \resizebox*{6.7cm}{6.7cm}{\includegraphics{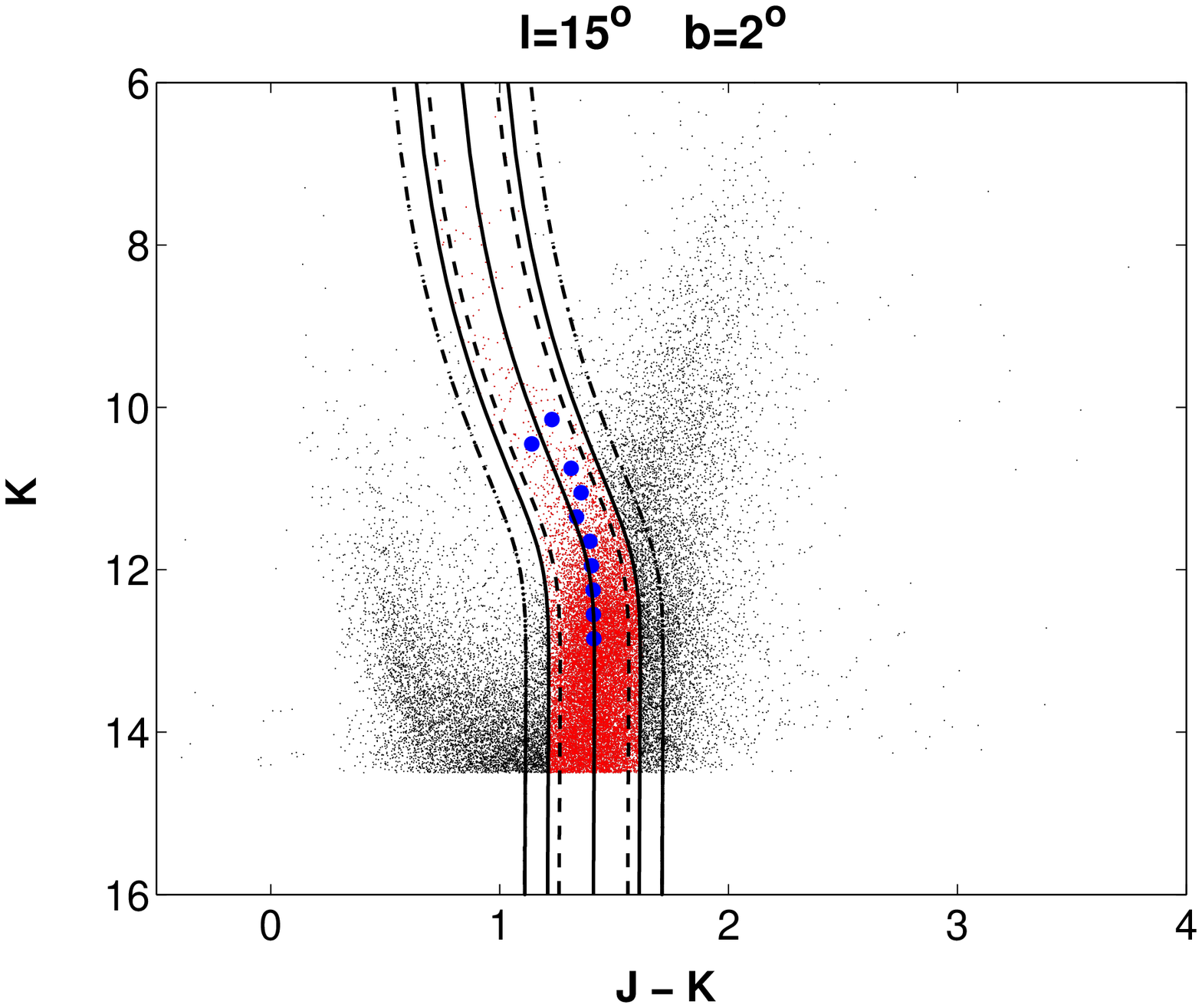}}
\resizebox*{6.7cm}{6.7cm}{\includegraphics{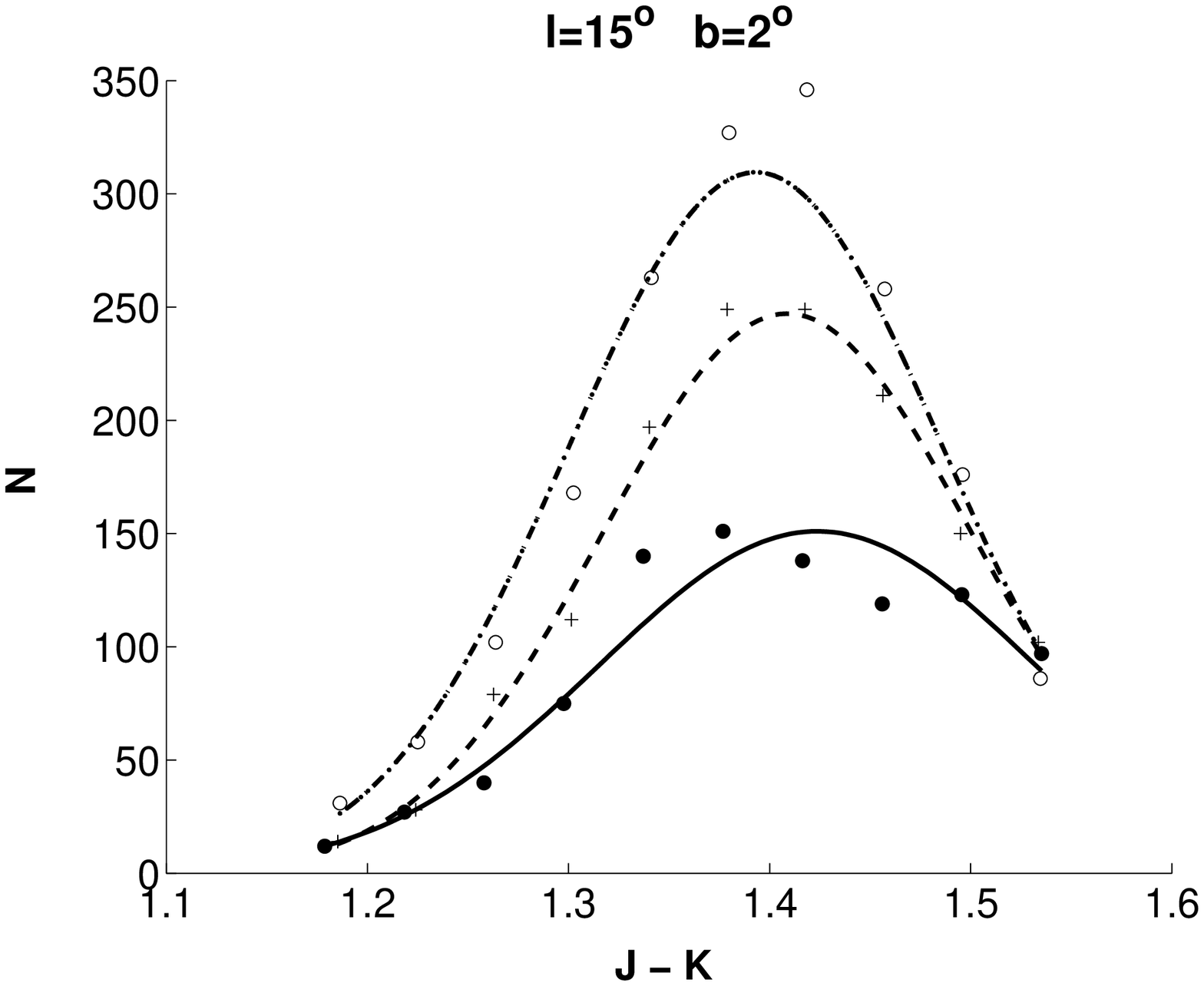}}\par}
\caption{Top: Colour--magnitude diagram for the line of sight $l=15^\circ $,
$b=2^\circ $ used in the present section with 2MASS data. 
The solid line shows the fitted trace that we assign 
to the  red clump giant population 
and the limits for the red clump giants within a 
width of 0.4 mag. Dashed lines show the limits for a width of 0.3 mag.
Dot--dashed line show the limits for a width of 0.6 mag.
Bottom: histograms of counts (per unit colour) 
for three fixed apparent magnitudes
and the corresponding best Gaussian fits: $m_K$=11.95 (dots/solid line), 
$m_K$=12.25 (crosses/dashed line), $m_K$ = 12.55 circles/dot--dashed line.
$\Delta m = 0.3$.}
\label{Fig:l15b2CM}

\end{figure}

\begin{figure}[!h]
{\par\centering \resizebox*{6.7cm}{6.7cm}{\includegraphics{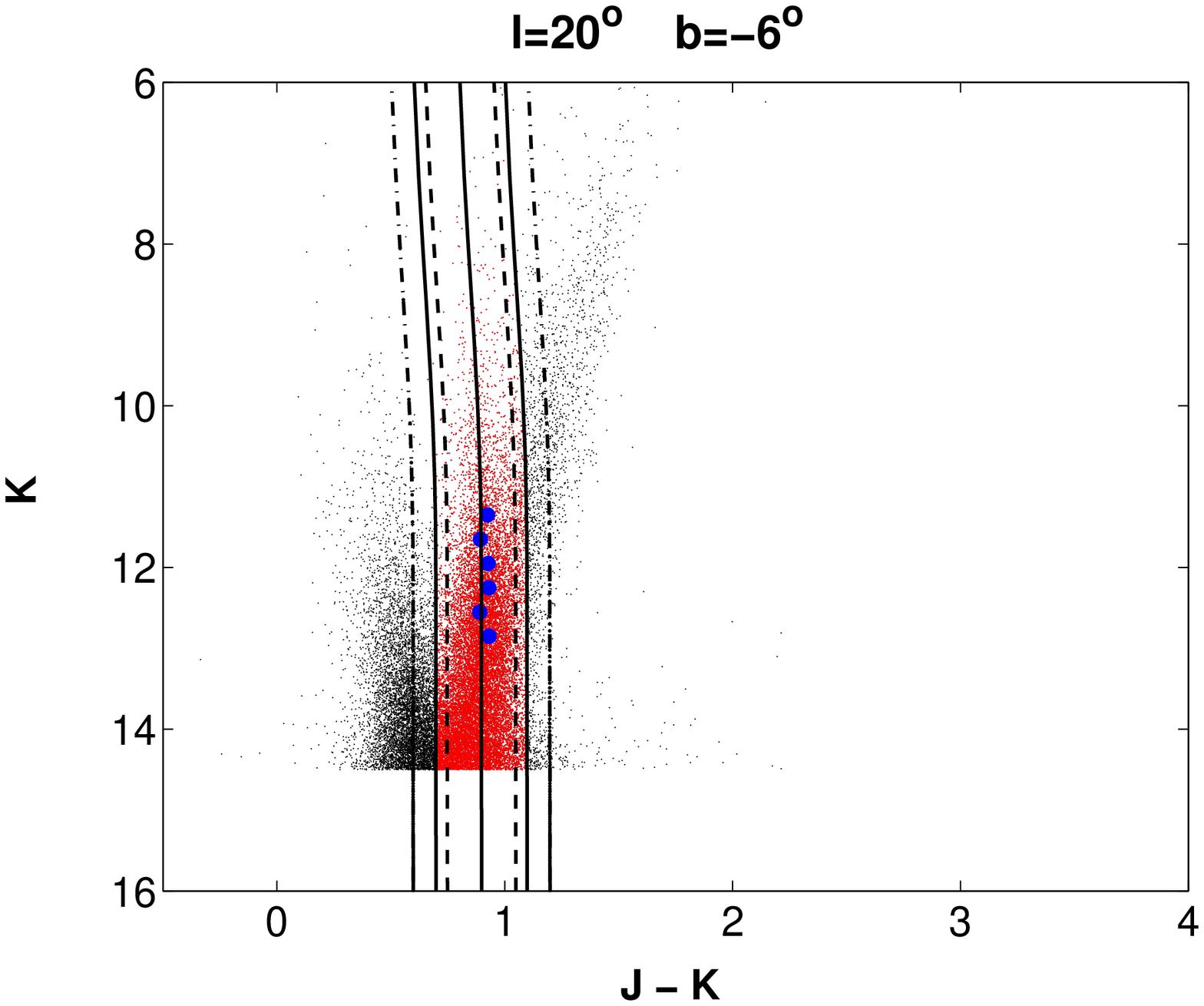}}
\resizebox*{6.7cm}{6.7cm}{\includegraphics{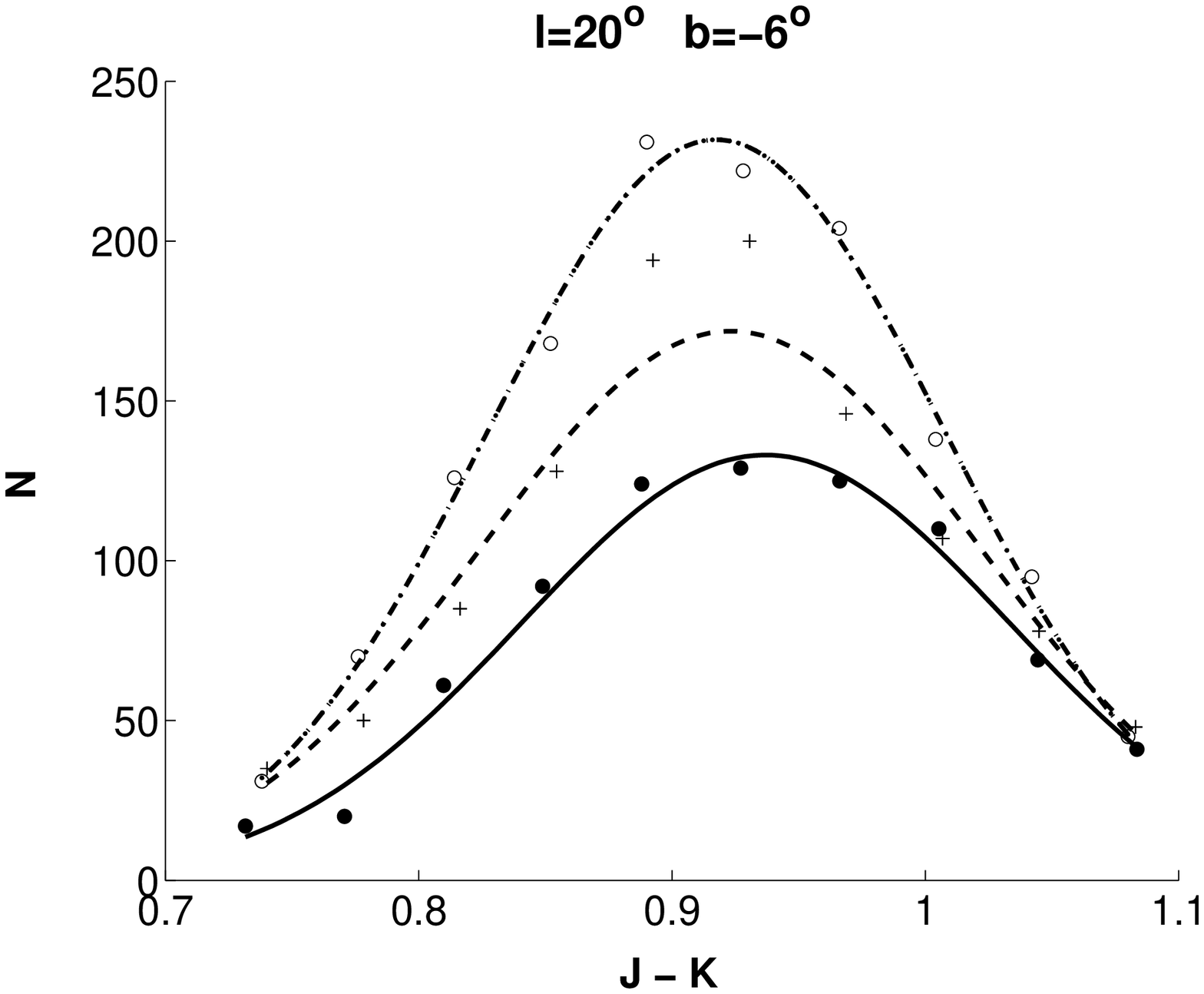}}\par}
\caption{Top: Colour--magnitude diagram for the line of sight $l=20^\circ $,
$b=-6^\circ $ used in the present section with 2MASS data. 
The solid line shows the fitted trace that we assign 
to the red clump giant population 
and the limits for the red clump giants within a 
width of 0.4 mag. Dashed lines show the limits for a width of 0.3 mag.
Dot--dashed line show the limits for a width of 0.6 mag.
Bottom: histograms of counts (per unit colour) 
for three fixed apparent magnitudes
and the corresponding best Gaussian fits: $m_K$ = 11.95 (dots/solid line), 
$m_K$ = 12.25 (crosses/dashed line), $m_K$=12.55 circles/dot--dashed line.
$\Delta m=0.3$.}
\label{Fig:l20v6CM}
\end{figure}

In L02 (\S3), some of us developed a method to
obtain the star density along a line of sight $(l,b)$ from an
analysis of near-infrared CMDs, such as
$m_K$ vs. $(J-K)$, which are the colours to be used here.
First, the trace produced in the CMD by the red clump giants (spectral
type: K2\,III) is identified. This is easy because they are much more abundant 
than the other giants (Cohen et al.\ 2000; Hammersley et al.\ 2000). Second, we count the number of stars
within a 0.4 mag width trace as a function of the apparent magnitude 
(examples are given in L02 and in Figs \ref{Fig:l15b2CM}, \ref{Fig:l20v6CM} 
of this paper). Third, because we know their absolute magnitude and 
intrinsic colour [$M_K=-1.65$, $(J-K)_0=0.75$; L02], we can obtain
the extinction and density as a function of  distance
along the line of sight. The correction of the mean extinction is 
carried out since we can measure the reddening of the K2\,III stars.
The details are explained in L02(\S 3), so they will not be repeated here.

Recent results in open clusters also give similar results for the intrinsic
luminosity of the red clump (Grocholski \&
Sarajedini 2002): magnitude $-$1.62 with a standard 
deviation of 0.21 and colour $(J-K)_0$ around 0.7 for the age of the old disc 
population with a small dispersion due to metallicity or age gradients. 
Pietrzy{\'n}ski et al. (2003) also show that the $K$-band
magnitude of red clump is a good distance indicator,
and that the colour has only a very slight dependence on the metallicity.
Theoretical predictions from isochrones (Salaris \& Girardi 2002)
agree with these values. Therefore,  counts can give
us information about the extinction along the line of sight as well as the
density of K2\,III stars: $\rho_{\rm K2\,III}(r,l,b)$, which leads to
$\rho_{\rm K2\,III}(R,z)$, assuming axisymmetry in the disc
($r$ = distance from the Sun, $R$ = distance from the centre of the Galaxy).

Concerning the completeness in crowded fields, in the worst
 cases ($l=15^\circ $, $b=2^\circ $) the survey is 
complete  to magnitudes $m_K\approx 13.2$ and $m_J\approx 14.4$. 
The reddening due to the extinction is less than $\Delta(J-K)=0.75$.
Since we use the red clumps with $(J-K)_0$ between 0.55 and 0.95
(the width of the strip is 0.4 mag), we will adopt a limit of $m_K<12.7$
(instead of 13.0 used in L02) to be sure of the completeness of the
sources.

Poisson errors will be important
for very bright stars (i.e. close stars, since the absolute magnitude
is constant). If we look towards the centre of the Galaxy, the Poisson
error will be important for distances from the Sun $r<3$ kpc (galactocentric
distance $R \ga 5$ kpc), but it will be reasonably
small for $R=2$--5 kpc, where the inner disc can be examined.
The method works reasonably well for $r$ 
greater than $\sim 3$ kpc, and this was tested with real data by L02,
who demonstrated that the parameters of the outer disc obtained by this
method are consistent with those obtained by other methods.
The global systematic errors of this method are less than  10\% 
for the regions with low extinction in a typical anticentre direction (L02).
In the present paper, the  areas are somewhat more extinguished
(up to 0.45 mag in $K$ instead of 0.2 mag in L02,  total extinction 
out to 7--8 kpc along the line of sight) and the regions
are towards the centre of the Galaxy instead of the outer disc regions used
in L02, which might give a somewhat higher contamination of M-giants.
This, however, will give systematic errors of the same order, as will be
shown below, perhaps only slightly larger, 
and the method is therefore appropriate for measuring the density of 
the inner disc stars.

\subsubsection{A simulation to test the method}
\label{.simulation}

To show how well the method works, we will carry out a Monte Carlo
simulation. We will generate a synthetic colour--magnitude diagram 
with a population model and a predescribed density model. From this CMD,
we will obtain the density along the line of sight using the 
method of extracting the clump giants, and this will be compared with
the original density.

The information about the absolute magnitudes and relative abundances
of all possible stellar spectral types was taken from the 
updated SKY model (Wainscoat et al.\ 1992; M. Cohen, private
communication). The red clump with the range of intrinsic colours of 0.2 mag
and range of absolute magnitudes of 0.3 mag (L02) is represented
 by the K0--5\,III population
with a peak in K2--3\,III; this population represents the clump with mean absolute
magnitude $-$1.65 and color $(J-K)_0=0.75$, with Gaussian dispersions
with the referred sigmas.
Poisson noise in the density is also introduced in the simulation.

\begin{figure}[!h]
{\par\centering \resizebox*{4.4cm}{4.4cm}{\includegraphics{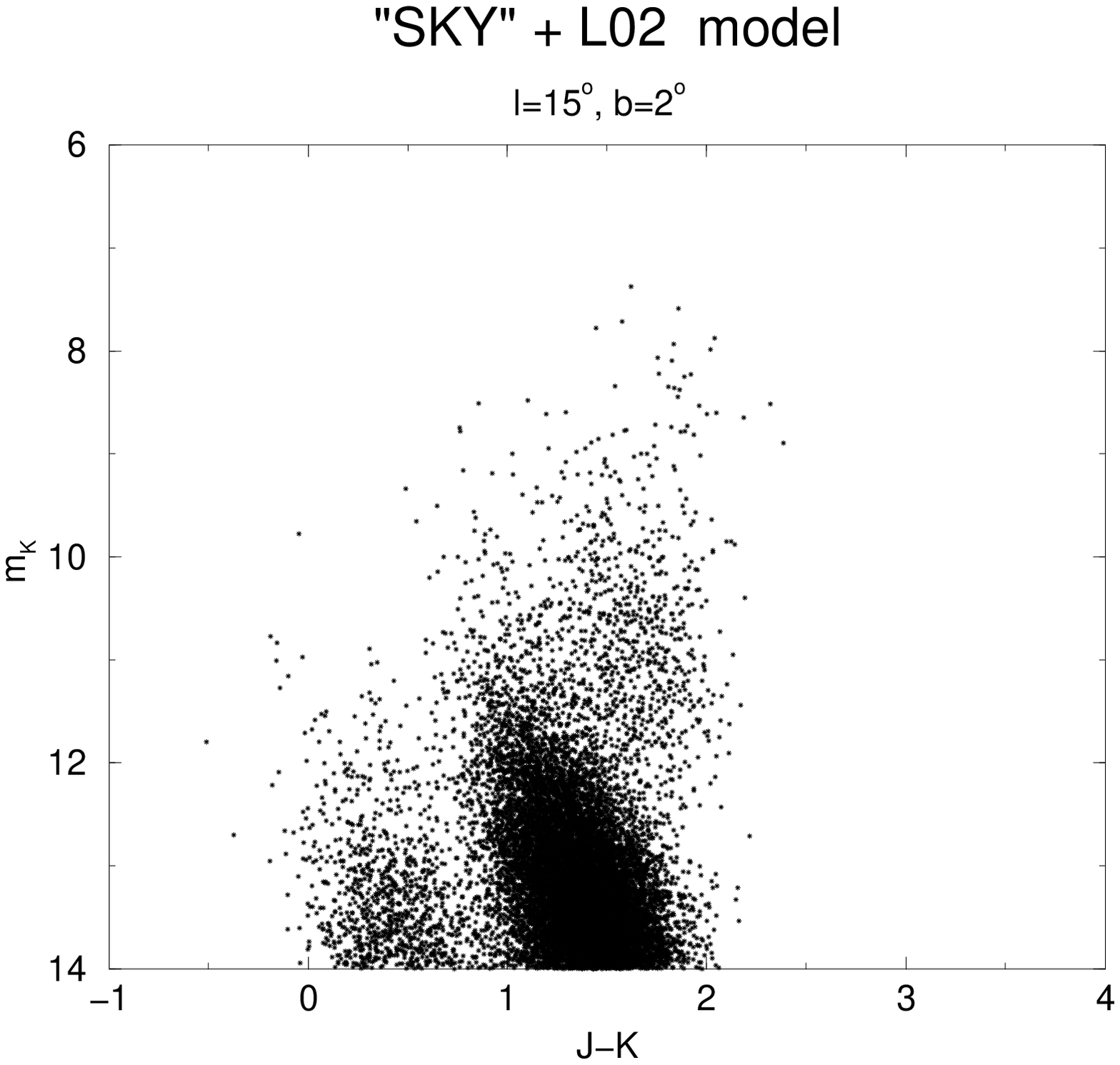}}
\resizebox*{4.4cm}{4.4cm}{\includegraphics{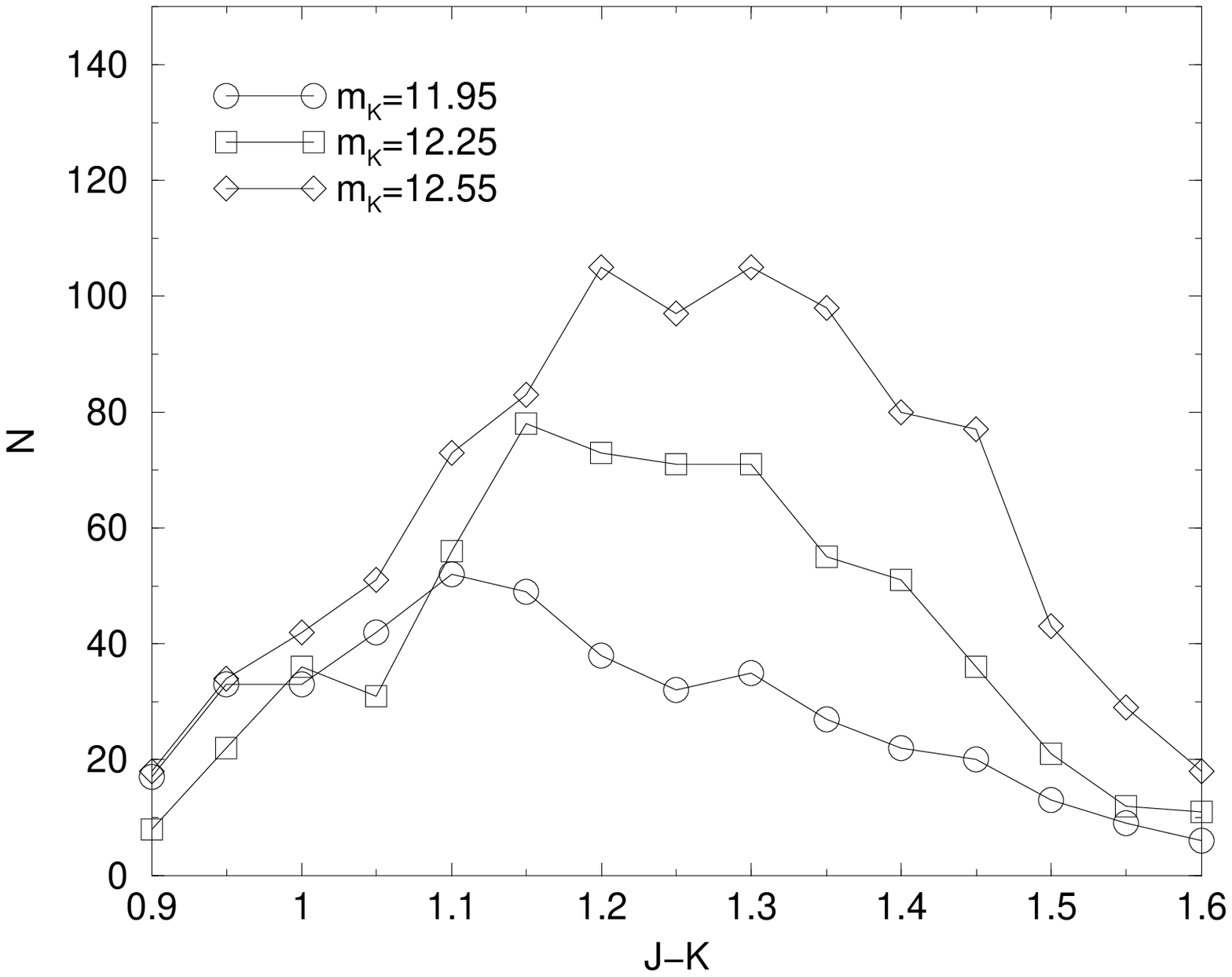}}
\resizebox*{4.4cm}{4.4cm}{\includegraphics{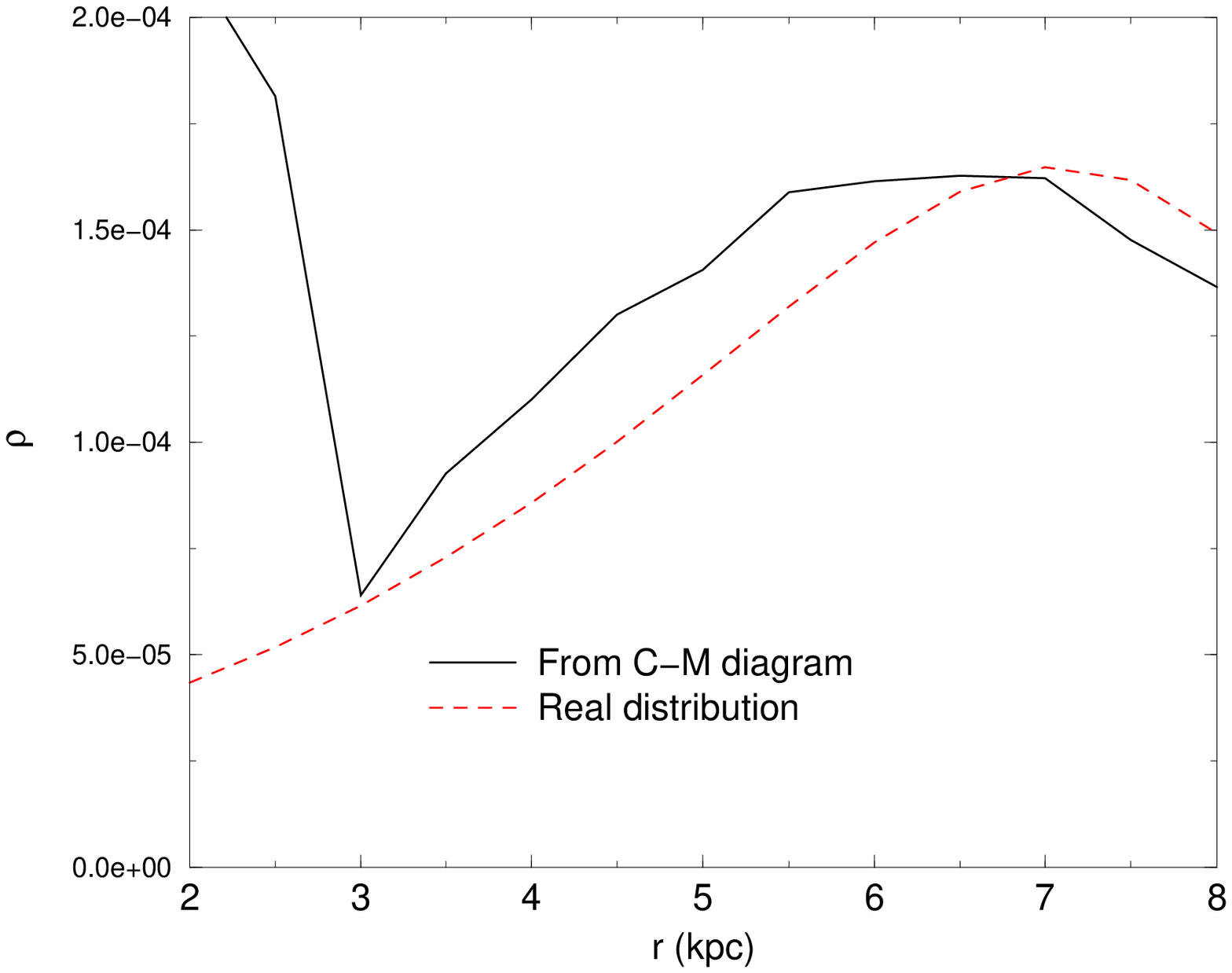}}\par}
\caption{Top: Synthetic colour--magnitude diagram derived from the SKY model
for the populations and relative abundance and an exponential disc
(larger density for low values of $R$). Centre: histograms of counts
for fixed $m_K$ ($\Delta m_K=0.3$). 
Bottom: density along the line of sight in the field $l=15^\circ $, 
$b=2^\circ $: comparison between the ``real'' assumed density and the
corresponding densities  obtained from the CMD analysis method with
width  = 0.4 mag of the synthetic colour--magnitude diagram.}
\label{Fig:CMmodel1}
\end{figure}

\begin{figure}[!h]
{\par\centering \resizebox*{4.4cm}{4.4cm}{\includegraphics{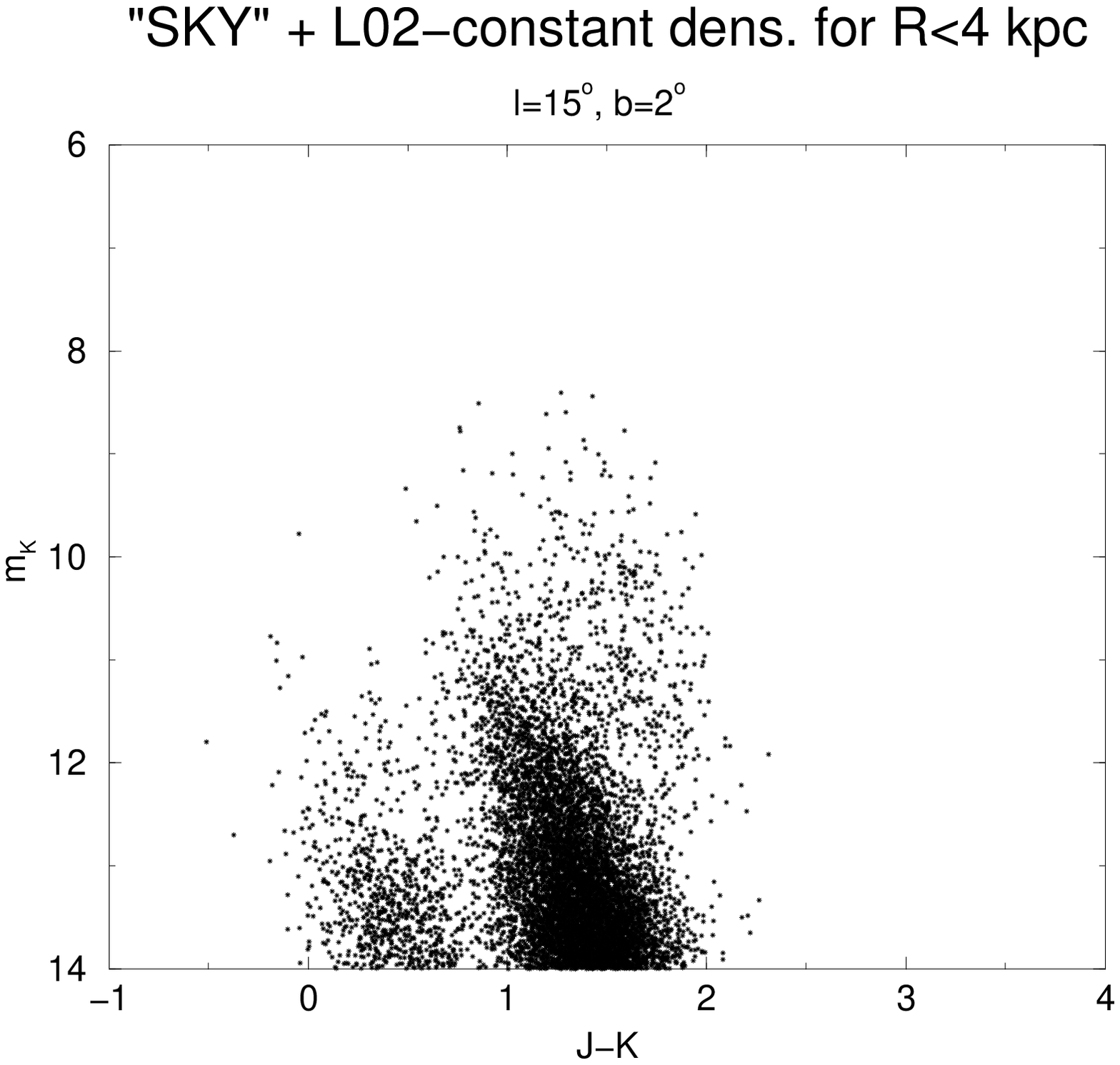}}
\resizebox*{4.4cm}{4.4cm}{\includegraphics{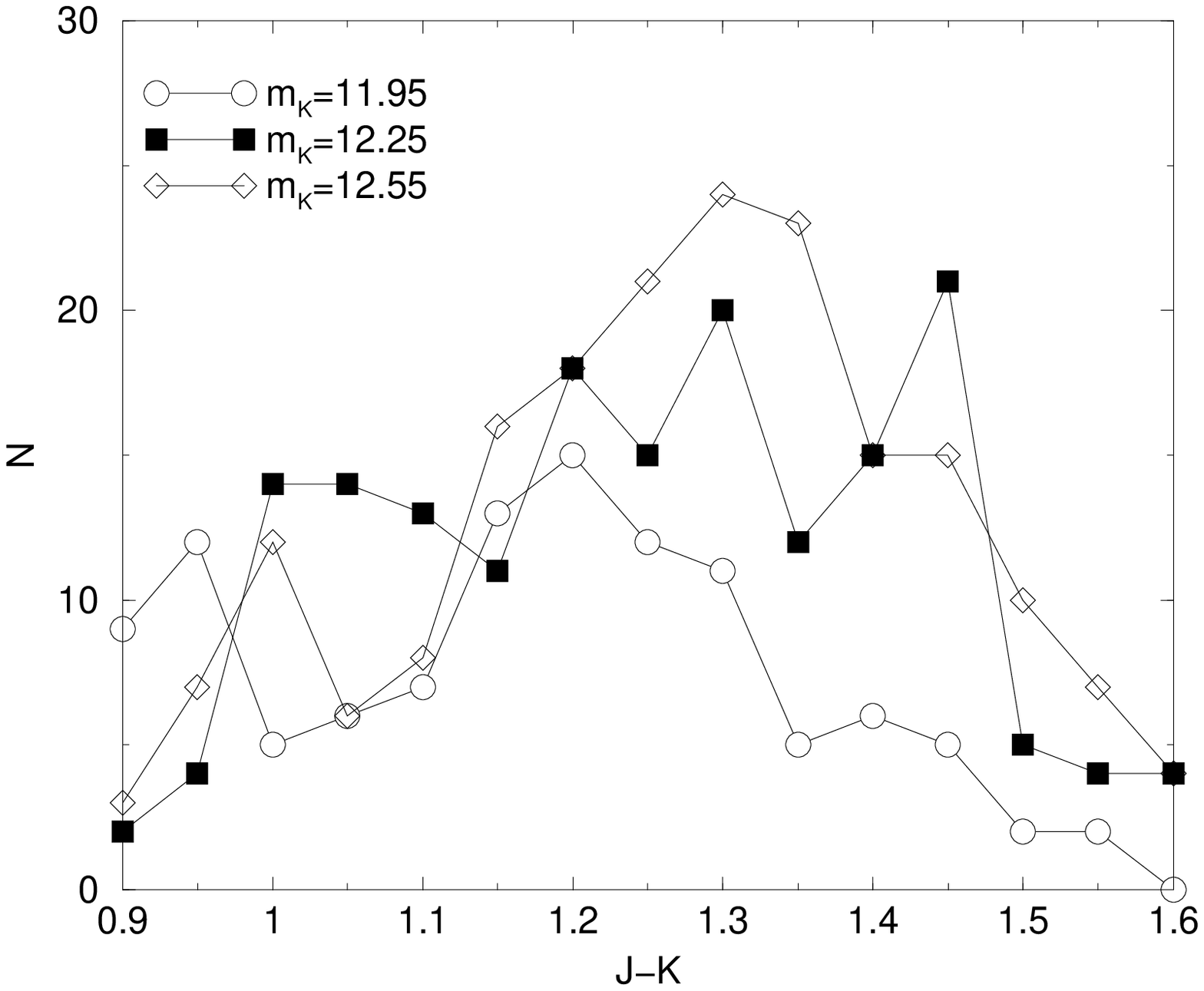}}
\resizebox*{4.4cm}{4.4cm}{\includegraphics{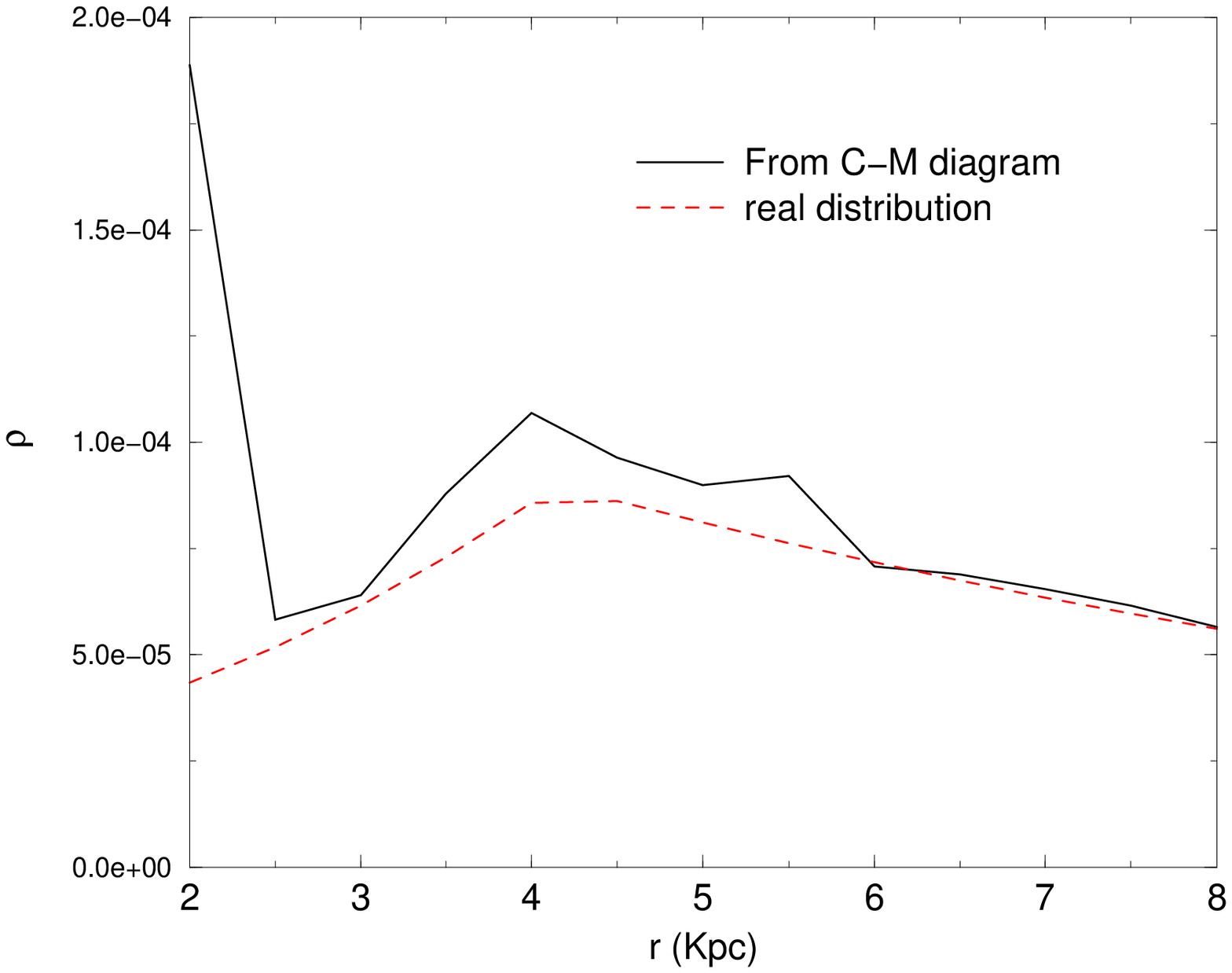}}\par}
\caption{Top: Synthetic colour--magnitude diagram derived from the SKY model
for the populations and relative abundance, and an exponential disc for $R>4$
kpc and constant density (independent of $R$) for $R<4$ kpc.
Centre: histograms of counts for fixed $m_K$ ($\Delta m_K=0.3$). 
Bottom: density along the line of sight in the field $l=15^\circ $, 
$b=2^\circ $: comparison between the ``real'' assumed density and the
corresponding  densities obtained from the CMD analysis method with
width = 0.4 mag of the synthetic colour--magnitude diagram.}
\label{Fig:CMmodel2}
\end{figure}

Since the region at $l=15^\circ$, $b=2^\circ $ has a higher
contamination of M-giants (because the ratio of distant/nearby stars is larger)
and the highest extinction, we will carry out our experiments
in this line of sight. This  is the one with the largest errors,
so the other lines of sight will have lower errors than the present one.
We carry out two experiments: 1) with a pure exponential disc: $\rho (R,z)
\propto e^{-R/H}e^{-|z|/h_z}$ ($R_\odot=7.9$ kpc, 
scalelength $H=0.25R_\odot$, scaleheight
$h_z=0.036R_\odot$; L02); 2) for a model with the same density for 
$R>4$ kpc, but $\rho (R,z)\propto 
e^{-|z|/h_z}$ (i.e.\ independent of galactocentric distance) for 
$R\le 4$ kpc. The results are shown in Figs \ref{Fig:CMmodel1} and
\ref{Fig:CMmodel2}.

The colour--magnitude diagrams in Figs \ref{Fig:CMmodel1} and
\ref{Fig:CMmodel2}  approximately reproduce the one observed in
Fig. \ref{Fig:l15b2CM}. They are not perfectly equal, but it is
not our purpose to discuss the validity of the SKY model here.
Our purpose  is to test the density extracted through the clump 
giant extraction method. This is tested in the right-hand part
of Figs \ref{Fig:CMmodel1} and \ref{Fig:CMmodel2} (for $r>3$ kpc; for lower
distances, the Poisson error and the M-giant contamination dominate,
but we are not interested in this range). In both cases,
the density is approximately that of the model. There are
 excesses of 20--30\% between $r$ = 4--6 kpc due the contamination
of the M-giants. The density distributions
in both cases can be clearly distinguished and the method is able
to detect the differences of the density in Figs \ref{Fig:CMmodel1}
and \ref{Fig:CMmodel2}.
This is the worst  case among the
selected lines of sight. Statistically
(when we use the data of all the lines of sight), the errors are
lower: roughly 10--15\% for $r$ = 4--6 kpc and even less beyond.  

\subsubsection{Uncertainties due to the patchiness of extinction}
\label{.extpat}

The extinction is corrected on average because we know the
reddening. In  extracting
the red clumps from the colour--magnitude diagram, we ask whether
the patchiness of the extinction, i.e.\ the
fluctuations in $A_K$ relative to the mean extinction, could
have some effect (characterized by dispersion $\sigma _{A_K}$). This is an
additional effect to consider which was not discussed in \S \ref{.simulation}.
It would produce an extra dispersion in the  apparent
magnitudes (which, if less than $\sim 0.3$ mag, produces negligible errors
in an exponential distribution; see L02, \S 3.3.1) and an extra dispersion
in the reddening, which can  result in the loss of some stars because they
move to a position in the CM diagram that is outside the selected
trace of red clump giants with width 0.4 mag (see L02, \S 3).
Since the intrinsic dispersion of colours is around 0.2 mag, the 
extra dispersion with respect to the central position of the trace
would be (for $\Delta (J-K)=1.52A_K$; Rieke \& Lebovsky 1985):

\begin{equation}
\sigma _{(J-K), {\rm centre\ trace}}=1.52\sigma _{A_K}
-\frac{\sigma _{A_K}}{\left(\frac{dm_K}{d(J-K)}\right)}
\label{extrared}
,\end{equation}
where the first term is the absolute value of the reddening
and the second term is the $(J-K)$ variation of the centre of the trace
corresponding to this extra extinction. $\frac{dm_K}{d(J-K)}$ is
the equation of the centre of the trace. From the relation between
apparent magnitude and the extinction, we get

\begin{equation}
\frac{dm_K}{d(J-K)}=\frac{\frac{dm_K}{dr}}{\frac{d(J-K)}{dr}}=
\frac{\frac{5\log_{10}e}{r}+\frac{dA_K}{dr}}{1.52\frac{dA_K}{dr}}
=0.658+\frac{1.43}{r\frac{dA_K(r)}{dr}}
\end{equation}

\begin{figure}
\begin{center}
\mbox{\epsfig{file=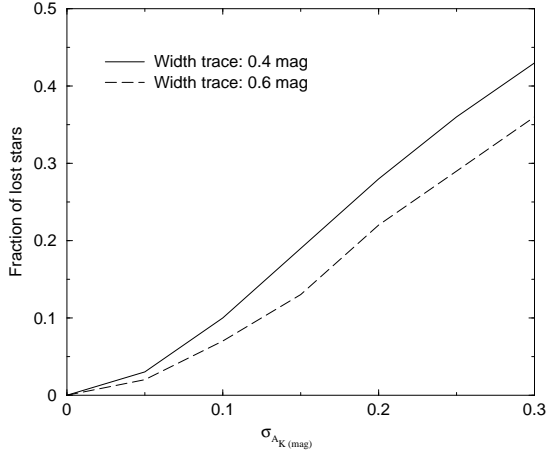,height=6cm}}
\end{center}
\caption{Fraction of lost stars due to the spread of the extinction
in the field $l=15^\circ $, $b=2^\circ $ for $r=5$ kpc.
Two values for the width of the K2\,III strip are used. In the subsequent 
analysis, a width of 0.4 mag is used.}
\label{Fig:lost}
\end{figure}

Consequently, for instance for $r=5$ kpc at $l=15^\circ $,
$b=2^\circ $ (corresponding to $R=3.4$ kpc): 
$\sigma _{(J-K), {\rm centre\ trace}}=1.13\sigma _{A_K}$.
For larger values of $r$ (lower $R$) the effect is somewhat lower
because the second term of eq. (\ref{extrared}) becomes more important:
for instance, $\sigma _{(J-K), {\rm centre\ trace}}=0.92\sigma _{A_K}$ 
(for $r=7$ kpc). This extra broadening of the distribution
will lead to a loss of stars for a fixed width of the trace.
Assuming that this extra broadening and the intrinsic broadening of the
distribution are Gaussian, the ratio of lost stars would be that
plotted in Fig. \ref{Fig:lost} for $l=15^\circ$, $b=2^\circ $. 

\begin{figure}[!h]
{\par\centering \resizebox*{6.7cm}{6.7cm}{\includegraphics{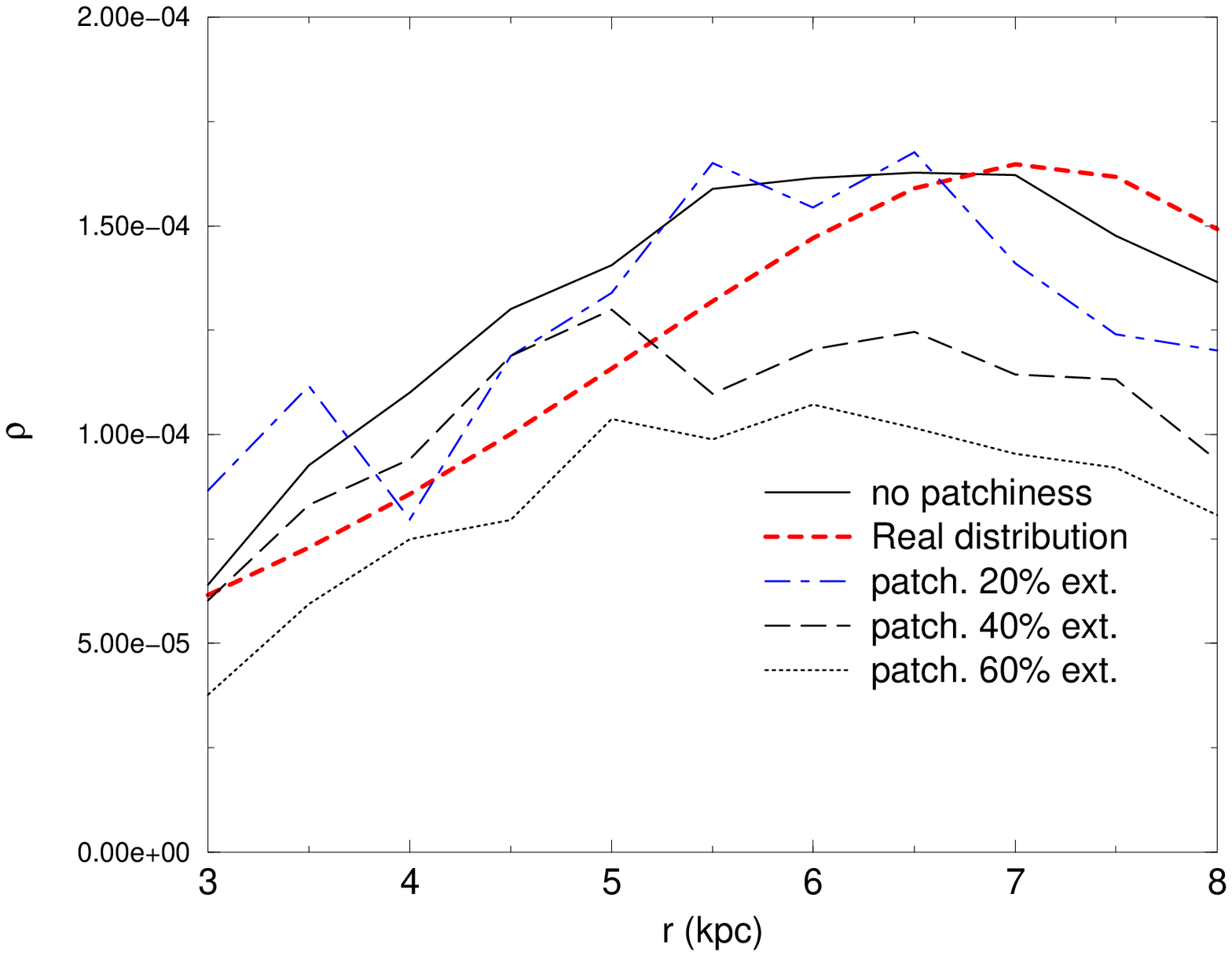}}
\resizebox*{6.7cm}{6.7cm}{\includegraphics{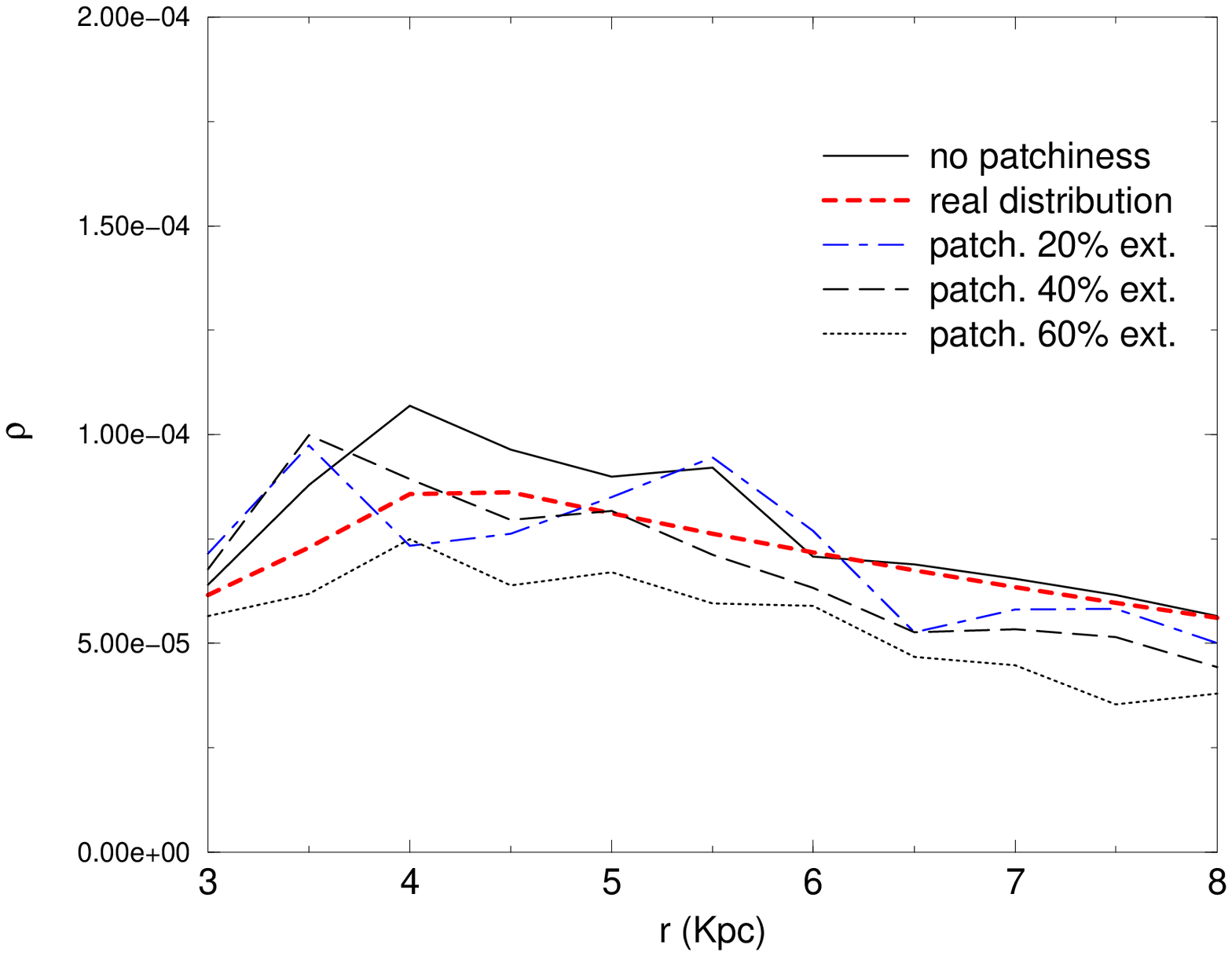}}\par}
\caption{Density along the line of sight in the field $l=15^\circ $, 
$b=2^\circ $ [a) exponential disc as in Fig. \ref{Fig:CMmodel1}; 
b) with constant density for $R<4$ kpc as in Fig. \ref{Fig:CMmodel2}]: 
comparison between the inputs of the density and the
density distribution obtained from the CMD analysis method with
width = 0.4 mag of the colour--magnitude diagram generated 
by a Monte Carlo simulation. Models with four different patchiness are plotted:
$\sigma _{A_K}=0,0.2A_K,0.4A_K,0.6A_K$.}
\label{Fig:ro1}
\end{figure}

A Monte Carlo simulation was also carried for the cases of Figs
\ref{Fig:CMmodel1} and \ref{Fig:CMmodel2} but adding the effect of 
patchiness of extinction. Fig. \ref{Fig:ro1} shows the result.
The effect of the patchiness of extinction might be important, but
only in cases with $\sigma _{A_K} = 0.6A_K$ or larger, while the
expected cases are under $0.2A_K$, as will be shown now.

To determine whether the patchiness
of extinction is significant it is necessary to evaluate the dispersion
of this extinction.
An examination of the Schultheis et al.\ (1999) and L01 
maps of extinction (with a resolution of 2 arcminutes) for $|l|<20^\circ$,
$|b|<2^\circ$ in $A_V$ gives a spread (due to both
patchiness and gradient) of around 20\% with respect to the mean
extinction, almost independently of which region in the centre of the Galaxy
is considered.

Lada et al. (1999) have shown that the variations in extinction through
the dark cloud IC 5146 for each background star 
due to small-scale fluctuations (once the large--scale gradients are 
removed, i.e. taking small regions) 
are less than 18\% of the mean extinction. If this is correct for the
Milky Way extinction in general, it would indicate   
that the patchiness is limited within the value of $\sigma _{A_K}=0.20A_K$, or even 
less if we integrate along a line of sight that passes through several
clouds, as would be the case for  near-plane regions. 

Since our region has 0.5 mag of mean extinction, 
this would give  $\sigma _{A_K}\approx 0.10$ mag, i.e.\ a loss of 10\%
of the stars in our case (7\% for width 0.6 mag).
This would be nearly the same for any $r$ between 
5 and 6.5 kpc ($R$ between 2.3 and 3.5 kpc). 
This is the worst of the cases, because the rest of the lines of
sight are less extinguished and  are consequently less patchy.
A model with a substantial deficit of stars (such as the one we will obtain below) 
gives an absence of 60--70\% of stars
with respect to the purely exponential L02 model at this latitude. 
A $\sigma _{A_K}$ value of around 0.40 mag, four times
our estimate, would be necessary to explain this,
which makes this method able to detect a deficit of stars of this order.

The effect of the gradient of extinction may be shown to be 
insignificant overall.
For the most extinguished region ($l=15^\circ $, $b=2^\circ $),
the effect of the gradient of extinction due to variation of $b$
between $1.5^\circ$ and $2.5^\circ$ is $\frac{\sigma _{A_K}}{A_K}=
\frac{\Delta b}{2\sqrt{3}}
\frac{d(ln\ A_K)}{db}$ and $\frac{d(ln\ A_K)}{db}\approx 0.49$
(Wainscoat et al.\ 1992);  the gradient in extinction in the
latitude direction produces a dispersion of  $\sigma _{A_K}\approx 0.07$
mag, which is certainly not Gaussian but for a rough estimate can be
considered so. From Fig.\ \ref{Fig:lost},
we will see that this can produce a 5--7\%  loss of stars, which
may be considered unimportant. 

\subsection{Application of the method}

We selected eight regions (see Table \ref{Tab:giants}) in 
$|l|$ = 15--20$^\circ$. In this zone, the disc is far enough from
the bulge that the contribution of bulge stars in the 
extracted red clump giant strip is small, 
and it is close enough to the Galactic Centre
to  have lines of sight reaching
$R$ = 2--4 kpc, where the inner disc will be studied.

Different latitudes were selected so as to have a representation of different
heights, except in plane regions, where the extinction might be
excessive and/or very irregular in spatial distribution, and the hypothetical 
in-plane bar would presumably contaminate the counts (L01).

\begin{figure}
{\par\centering \resizebox*{8.5cm}{17cm}{\includegraphics{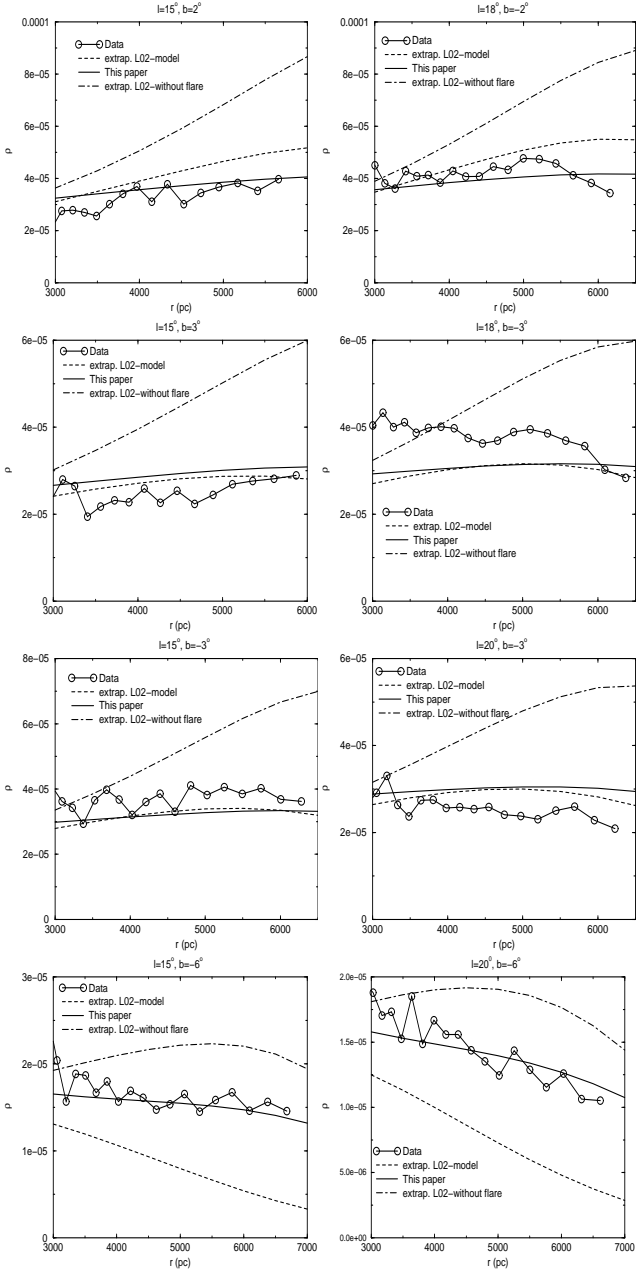}}\par}
\caption{Density along the different lines of sight.
An extrapolation of the exponential L02 model does not fit these data: 
the model gives only  $\sim$30\% of the density at $b=-6^\circ $, which means
that the extrapolation of the flare (the decrease in $h_z$ with decreasing $R$)
found in L02 for the external disc is not appropriate for the inner disc. 
To fit these data, a density nearly independent of $R$ must
be introduced (thicker solid line in the plot), which will be obtained in
this paper [eqs. (\ref{hz}) and (\ref{rogig})].}
\label{Fig:rho_d}
\end{figure}

The density $\rho_{\rm K2\,III}(R,z)$ along the eight lines of sight was obtained
(see Figs \ref{Fig:rho_d}).
An extrapolation of the exponential L02 model does not fit these data: 
the model gives only $\sim$30\% of the density at $b=-6^\circ $, which means
that the extrapolation of the flare (the decrease in $h_z$ with decreasing $R$)
found in L02 for the external disc is not appropriate for the inner disc. 
If we extrapolate the exponential L02 disc but without a flare, the result is 
better for $b=-6^\circ $, but there is an excess of $\sim$200\% in the 
density at $|b|=2^\circ $. It is clear that an exponential
disc does not work either with or without a flare of increasing h$_z$ for
increasing $R$, according to the extrapolation of L02, or 
with any intermediate option. 

\begin{figure*}
\begin{center}
{\par\centering \resizebox*{16cm}{13cm}{\includegraphics{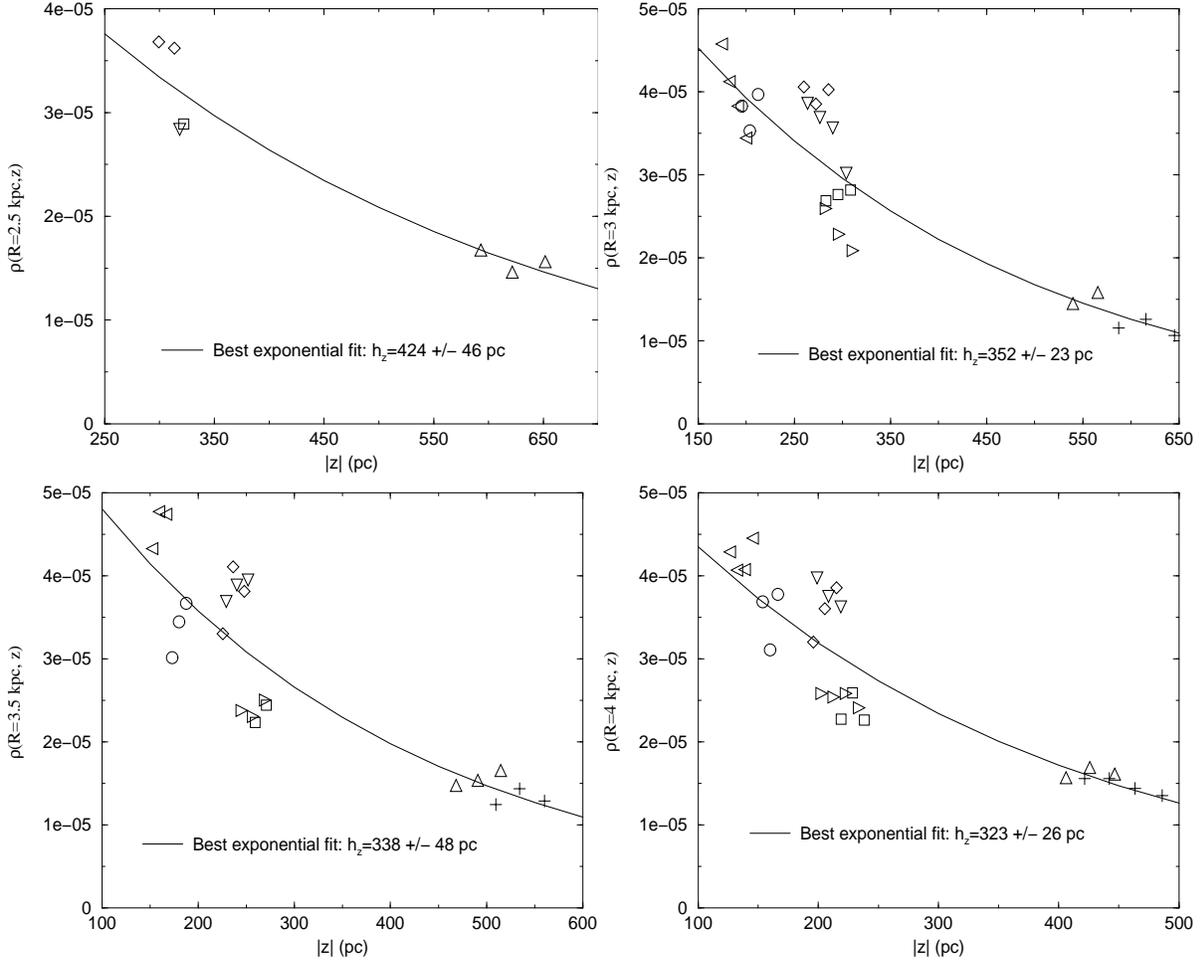}}\par}
\end{center}
\caption{Density of red clump giants, $\rho_{\rm K2\,III}(R,z)$,
at $R=$2.5, 3.0, 3.5 and 4.0 kpc. The intervals
of $R$ are in steps of 0.5 kpc, and the data in each diagram
belong to $R-0.25$ kpc $<$ galactocentric radius$ < R+0.25$ kpc.
Circles: line of sight $l=15^\circ $, $b=2^\circ $; squares:
$l=15^\circ $, $b=3^\circ $; diamonds: $l=15^\circ $, $b=-3^\circ $;
triangles up: $l=15^\circ $, $b=-6^\circ $; triangles left:
$l=18^\circ $, $b=-2^\circ $; triangle down: $l=18^\circ $, $b=-3^\circ $;
triangle right: $l=20^\circ $, $b=-3^\circ $;
plus: $l=20^\circ $, $b=-6^\circ $.
Best exponential fit vs. $z$ is also plotted.}
\label{Fig:R}
\end{figure*}

In Fig. \ref{Fig:R}, the densities are plotted as a function of $z$ for
different $R$ in $R<4$ kpc (note that, for the calculation of $z(r,l,b)$, we take
into account that the Sun is 15 pc above the plane, but not
the effects of the warp described in L02 since they are negligible for the inner disc; 
the warp is important for regions near $|l|=90^\circ $ and $R>10$ kpc). 
In these plots, we observe how the
scaleheight changes with $R$; it increases towards the centre:
$h_z(2.5 {\rm \ kpc})=424\pm 46$ pc,
$h_z(3 {\rm \ kpc})=352\pm 23$ pc, $h_z(3.5 {\rm \ kpc})=338\pm 48$ pc,
$h_z(4 {\rm \ kpc})=323\pm 26$ pc. A weighted fit of a linear law 
of $h_z$ gives:

\begin{equation}
h_z=317\pm 17-[R(\rm kpc)-4]48\pm 20 {\rm \ pc},
\label{hz}
\end{equation}
for 2.25 kpc $<$ R $<$ 4.25 kpc.
Therefore, we have a higher scaleheight
in the inner disc with the opposite trend
to that of the outer disc, which, according to L02, ranges from 230 and 285 pc between $R=6$ kpc and
$R_\odot$: here, $h_z$ slightly decreases with increasing $R$. 
Eq. (\ref{hz}) is valid for $R<4.25$ kpc while that given by L02 
is valid for $R>6$ kpc. In the range between 4 and 6 kpc
the scaleheight dependence on radius is somewhat uncertain but is
presumably the region where $h_z$ is at a minimum.

One might suspect  the contamination of the bulge to be
 responsible for this increase in the scaleheight. However,
the bulge contribution is negligible. Using the 
bulge model of axial ratios 1:0.54:0.33 of 
L\'opez-Corredoira et al.\ (2000), we find that the largest bulge/disc ratio
is for $l=15^\circ $, $b=-6^\circ$, $R=2245$ pc which gives a contribution
$\rho _{\rm bulge}= 5\times 10^{-3}$  pc$^{-3}$.
Since the K2 giants are $\approx 10^{-4}$
times the total number of stars (L02), this gives 
$\rho_{\rm K2\,III,bulge}\approx 5\times 10^{-7}$  pc$^{-3}$ = 5\%
of the total K2\,III density, which is less than the other errors. 
Higher ellipticity models of the triaxial bulge (such as that of B02) would give 
even less bulge contamination. This is the most favourable
line of sight for the bulge; for the other $(l,b,R)$ the
bulge contamination is even lower.

\begin{figure}
\begin{center}
\mbox{\epsfig{file=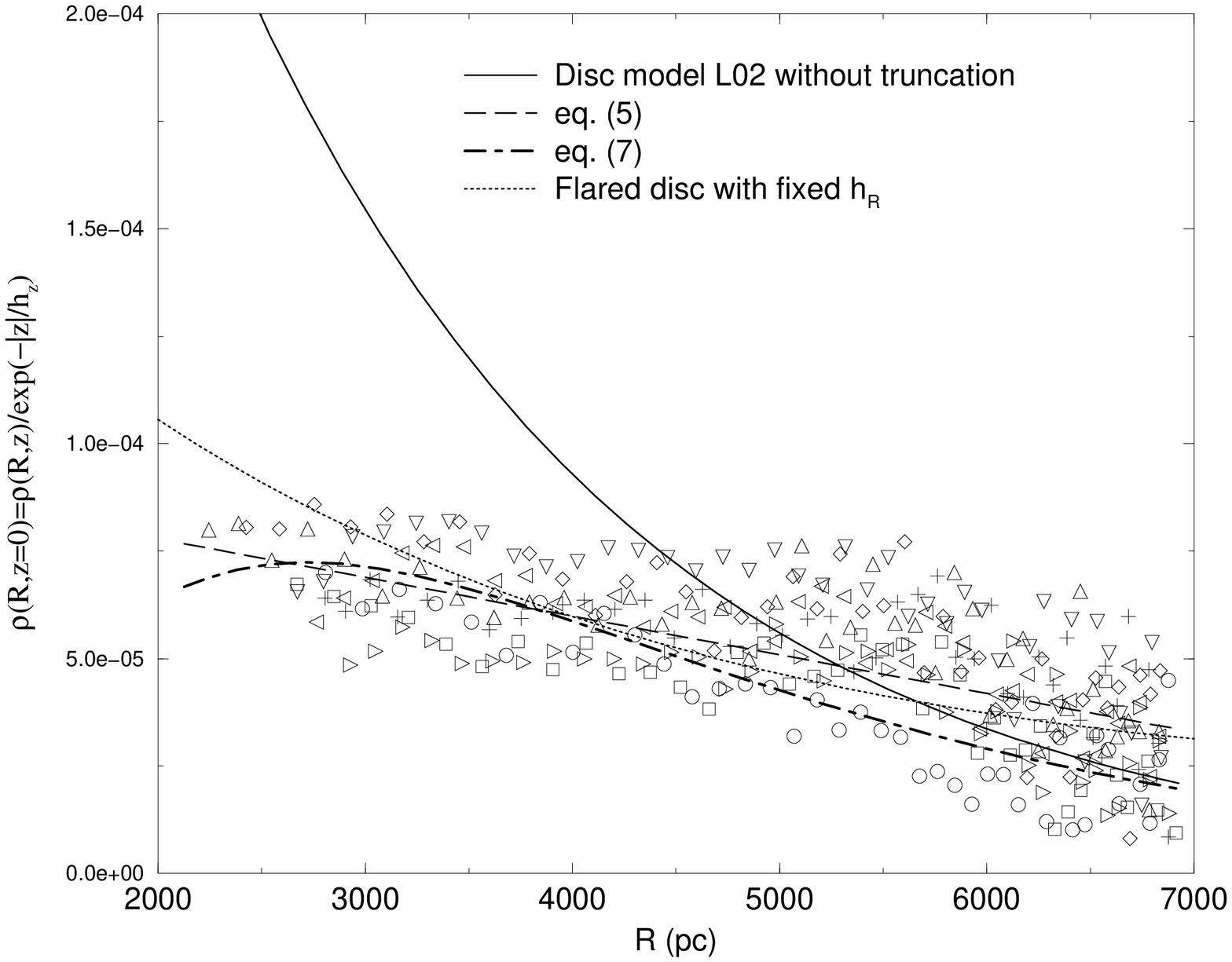,height=8.75cm,angle=270}}
\end{center}
\caption{Density of red clump giants in the plane [derived indirectly
from the counts, using the scaleheight $h_z(R)$ from eq. (\protect{\ref{hz}}) for $R<5.9$ kpc
and from L02 for $R>5.9$ kpc]. Note that the data for $R>5$ kpc ($r<3$ kpc)
have large errors due to Poisson noise and M-giant contamination.
Exponential disc model taken from an extrapolation of L02 outer 
disc model.
Circles: line of sight $l=15^\circ $, $b=2^\circ $; squares:
$l=15^\circ $, $b=3^\circ $; diamonds: $l=15^\circ $, $b=-3^\circ $;
triangles up: $l=15^\circ $, $b=-6^\circ $; triangles left:
$l=18^\circ $, $b=-2^\circ $; triangle down: $l=18^\circ $, $b=-3^\circ $;
triangle right: $l=20^\circ $, $b=-3^\circ $; plus: $l=20^\circ $, $b=-6^\circ $.}
\label{Fig:densplano}
\end{figure}

Once we know the scaleheight as a function of the radius, $h_z(R)$,
we can calculate

\begin{equation}
\rho_{\rm K2\,III}(R,z=0)\equiv \frac{\rho_{\rm K2\,III}(R,z)}{e^{\frac{-|z|}{h_z(R)}}}
,\end{equation}
where $h_z(R)$ is taken from eq.\ (\ref{hz}) for $R<5.9$ kpc and from L02
for $R>5.9$ kpc, in order to keep the continuity of $h_z$.
This is represented in Fig. \ref{Fig:densplano}. For $R>5$ kpc the
dispersion  is larger because the points  stem from the counts of very
bright red clump sources (which are few in number), so they have significant Poisson
noise and  also  an important contamination of M-giants. 
It is remarkable how low the dispersion of points is for
$R<4$ kpc and how clearly the density does not increase
exponentially towards the centre. If eq.\ (\ref{hz}), representative
of the scaleheight in the inner disc, was greatly in error, the dispersion for
$R<4$ kpc would be much higher, since we have used data
with $|b|$ between 2$^\circ$ and 6$^\circ$.

If we fit a linear relationship in  Fig. \ref{Fig:densplano} 
for $R<4.25$ kpc, we obtain:
 
\[
\rho_{\rm K2\,III}(R,0)=(4.57\pm0.33)[1
\]\begin{equation}
-0.150\pm0.033(R({\rm kpc})-4)]
\rho_{{\rm K2\,III},\odot};
\label{rogig0}
\end{equation}

for 2.25 kpc $<$ R $<$ 4.25 kpc, where $\rho_{{\rm K2\,III},\odot }=1.31\times 10^{-5}$ pc$^{-3}$ (L02).
This shallow rise is incompatible with an extrapolation of the outer exponential
disc towards the centre. 
Although the data to 2 kpc  show a slight increase of the density
towards the centre, a constant density is also consistent within
the systematic errors ($\Delta \rho /\rho \sim 20$\%).
If we neglect the small gradient with the galactocentric radius in the plane
(since the systematic errors are larger than these small variations
and we cannot be sure that the gradients are real), then

\begin{equation}
\rho_{\rm K2\,III}(R,z=0)\approx 5.1 \rho_{{\rm K2\,III},\odot}
\label{rogig}
,\end{equation}
for 2.25 kpc $<$ R $<$ 4.25 kpc.
We can also fit a more complex law, like a modified exponential
(already used for instance by L\'epine \& Leroy 2000): 

\begin{equation}
\rho (R,z=0)=\rho _\odot 
e^{\left(\frac{R_\odot}{H}+\frac{H_2}{R_\odot}\right)}
e^{-\left(\frac{R}{H}+\frac{H_2}{R}\right)}
\label{expdob0}
,\end{equation}
where $\rho _\odot=1.31\times 10^{-5}$ pc$^{-3}$ and $H=1.97$ kpc are 
already known from L02, and the parameter $H_2$ is fitted with our data
(see Fig. \ref{Fig:densplano}) with a best value of:

\begin{equation}
H_2=3740 \pm 130 ~\rm{pc}
\label{h2}
.\end{equation}

\begin{figure}[!h]
{\par\centering 
\epsfig{file=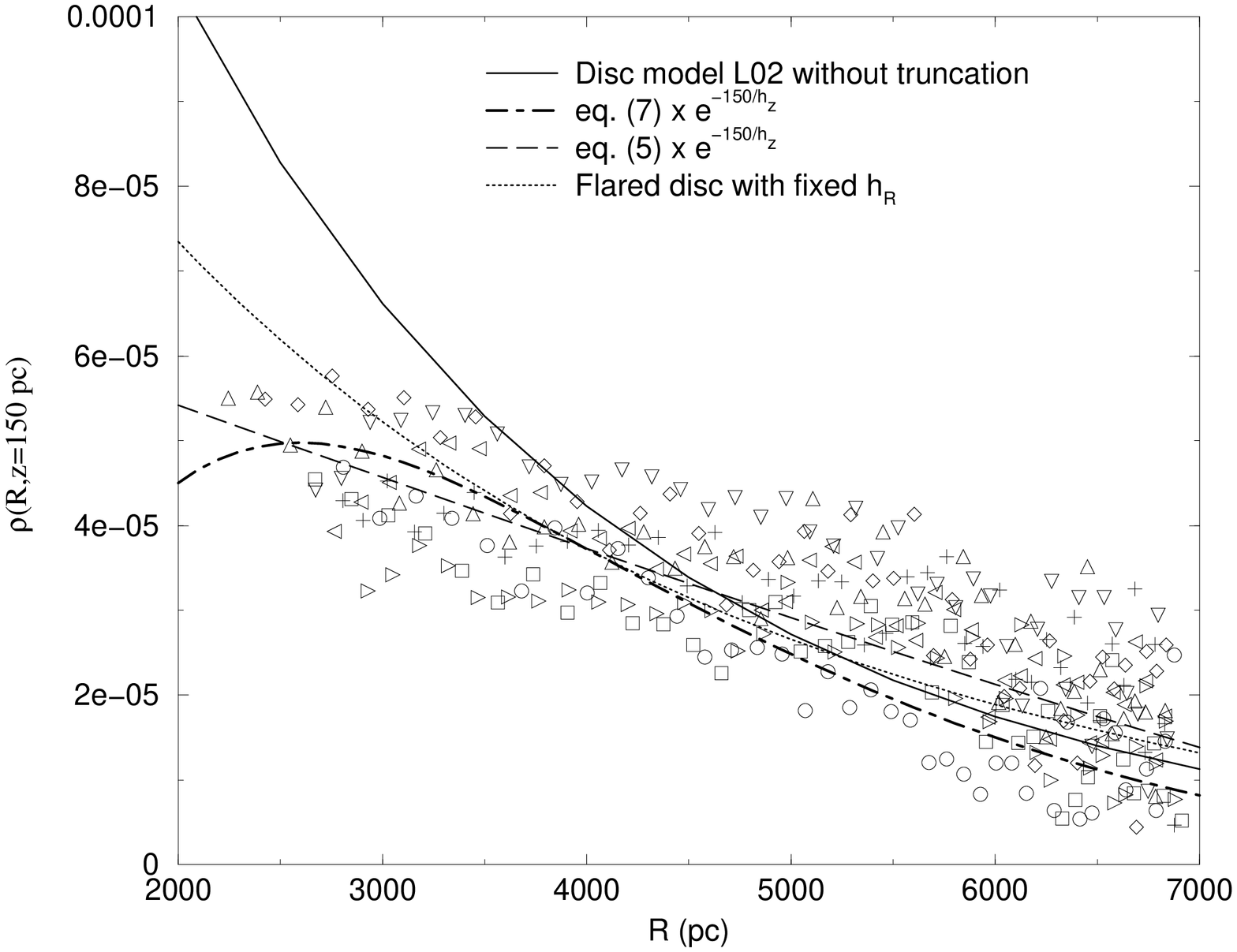,height=8.5cm,angle=270}
\epsfig{file=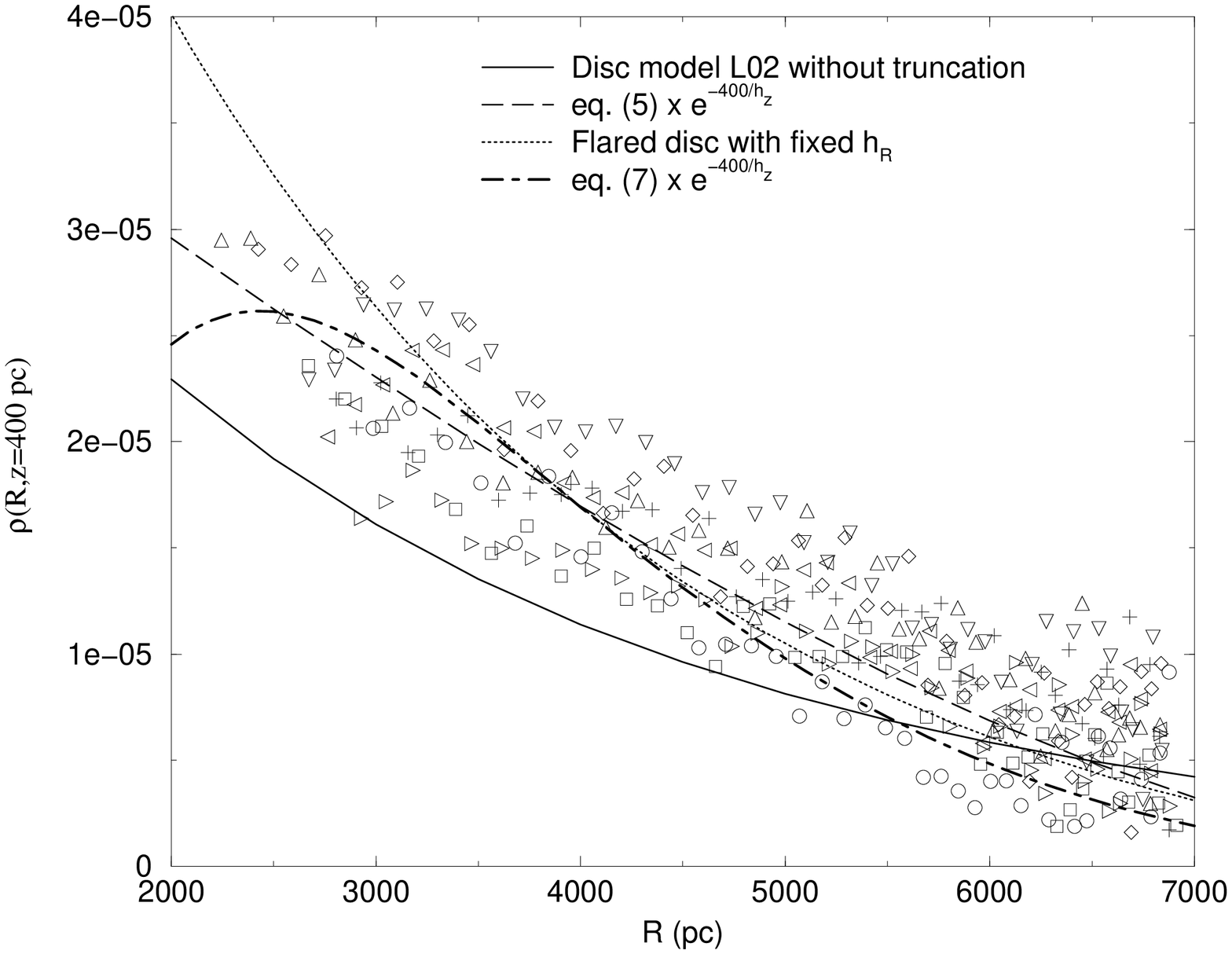,height=8.5cm,angle=270}\par}
\caption{Density of red clump giants at fixed $z$ = 150 pc and 400 pc 
[derived indirectly from the counts using the 
scaleheight $h_z(R)$ from eq. (\protect{\ref{hz}}) for $R<5.9$ kpc
and from L02 for $R>5.9$ kpc]. Symbols are the same than in Fig. \protect{\ref{Fig:densplano}}.}
\label{Fig:densz}
\end{figure}

The fit of eqs (\ref{rogig0}) and (\ref{expdob0}), or eq. (\ref{rogig}),
together with eq. (\ref{hz}),  is also applicable 
away from the plane, as can be observed in 
Fig. \ref{Fig:densz}. 
Observe that for lower $z$, the observed density is
lower than the extrapolation of the exponential law, while
for higher $z$ the density increases towards the centre more rapidly
than the extrapolation of the L02 model. That is, the deficit of
stars with respect to the L02 disc affects the regions near the plane.
Also, Fig.\ \ref{Fig:rho_d} shows that the present model fits 
 the density obtained directly from the counts (as a function 
of the distance from the Sun for the different lines of sight).

\subsubsection{Flare in the inner disc}
Another way to interpret the observed effect is the existence of a flare in the inner disc,
which redistributes the stars of the plane in off-plane regions. In L02 we considered
an exponential distribution for the density where the flare affects both  the scaleheight
(which increases with the galactocentric distance) and  the flare scalelength (which changes,
causing  the scalelength of the disc space density, H, to remain constant; see L02 for details). However, the
observed effect in the in-plane density  at the central Galaxy can also be reproduced by an inner flared disc. If we
consider a disc with a fixed exponential scalelength  of the disc surface density, and we
flare this disc vertically
by the flaring law defined in eq.\ (\ref{hz}), conserving the mass, i.e., the stellar surface
density, we find a radial law for
the disc density consistent with the observed counts. Considering a fixed scalelength of
$h_R$ = 2.4 kpc, for example, the 
agreement is quite good between
the observed and predicted densities in the plane in the range 2--8 kpc. As a comparison, in
L02 the scalelength of the disc surface density is $h_R$ = 3.2 kpc in the outermost disc
($R>$10 kpc), but the
flare scalelength changes as the intrinsic scalelength of the disc varies.

In Fig.\ \ref{Fig:densplano}, we  represent the
predicted densities for four different models: i) the extrapolation of the exponential L02 model, 
 ii) a linear relationship for
$R<4.25$ kpc as given by eq. (\ref{rogig0}), iii) the modified exponential given by eq. (\ref{expdob0}) and iv) a flared 
disc with a fixed scalelength of the surface density ($h_R$ = constant). In each case, 
except for the extrapolation of the L02
model, there is a good match between the predicted and  observed density. This 
agreement is also observed away from the plane, where models ii), iii), and iv)  predict
roughly the
same star density (Fig. \ref{Fig:densz}).

However, there are basic differences between a vertical flare and a ``hole'' in the
central disc, the latter being considered as a deficit in the stellar density. While the former
retains the vertical column  density, in star numbers, a hole in the sense defined here 
will decrease that column density, and will appear to an external observer looking at the
Galaxy face on as a real deficit in the stellar distribution of the central regions as
compared to the surroundings.

We plan to investigate further the real nature of this feature of the stellar
distribution in a subsequent paper. In the rest of this paper, we will refer to this
observed feature as a ``hole'', without  prejudging the true origin of this decrease in
the stellar distribution with respect to the exponential prediction.

\subsubsection{Calculations
with different band widths in the subtraction of
red clump giants}

The variation in the width
of the band to extract the red clumps could alter the results slightly
due to the introduction of certain systematic errors. In this
subsection, we will examine how large these effects are.

We have already seen (\S \ref{.simulation}) that the contamination of the 
giants cannot emulate such a strong deficit of stars. Furthermore, the contamination
would increase the counts with respect the number expected in the L02
model (and the normalization of this model was carried out
in regions where the contamination is very low) but never decrease them. 
A more important source of error could be the higher loss of stars 
for fainter magnitudes due to the spread away from the 0.4 mag width strip,
but we have already verified (\S \ref{.extpat}) that this effect is unimportant 
(less than 10\% according to calculations).

For a further test of the results,   
we have checked that changing the aperture
of the trace (which also changes the contamination of other sources:
dwarfs and M giants; see L02, \S 3) does not significantly  change 
the  density obtained for the inner disc.
We have derived expressions (\ref{hz}) and (\ref{rogig0}) for width
0.6 instead of 0.4  used above, with the result:

\begin{equation}
h_z=282\pm 9-[R(\rm kpc)-4]48\pm 15 {\rm \ pc};
\end{equation}
\[
\rho_{\rm K2\,III}(R,0)=(5.55\pm 0.34)[1
\]\[-0.024\pm 0.020(R({\rm kpc})-4)]
\rho_{{\rm K2\,III},\odot};
\]
\begin{equation}
{\rm \ \ for\ 2.25\ kpc<R<4.25\ kpc}
,\end{equation}

which is  compatible with the previous result.
If we reduce the width rather than increasing it, for instance, to a 0.3 mag width
instead of 0.4 mag, the results are:

\begin{equation}
h_z=325\pm 16-[R(\rm kpc)-4]49\pm 21 {\rm \ pc};
\end{equation}
\[
\rho_{\rm K2\,III}(R,0)=(5.15\pm 0.40)[1
\]\[-0.165\pm 0.037(R({\rm kpc})-4)]
\rho_{{\rm K2\,III},\odot};
\]

\begin{equation}
{\rm \ \ for\ 2.25\ kpc<R<4.25\ kpc.}
\end{equation}

There are some small differences that can stem from the different degree
of contamination from stellar types other than K2\,III 
(a width of 0.6 is more contaminated), and the different degree of patchiness
of extinction (this affects the width 0.3 more; see \S \ref{.extpat}).
These small differences do not change any of the conclusions. 

On the other hand,
the possible errors in the characterization of the K2\,III stars would move the position 
of the beginning of this non-exponential distribution, 
but never reproduce the shape of 
$\rho_{\rm K2\,III}(R,z=0)$ decreasing towards the centre. 


\subsubsection{Summary of the results of this section}

A rough expression for the central inner disc density is:

\begin{equation}
\rho (R,z)\approx 5.1\rho _\odot e^{\frac{-|z|}{(509-48R({\rm kpc})\ {\rm pc})}}
\label{hole}
\end{equation}
This is applicable to 2.25 kpc $< R <$ 4.25 kpc, and the region of transition
with the outer L02 disc would be abrupt. A more realistic model is likely to have
a smooth transition between 4 and 6 kpc, perhaps something like
eq. (\ref{expdob}). This expression, eq.\ (\ref{hole}), is equivalent to:

\begin{equation}
\rho (R,z) \approx \rho _\odot e^{\frac{-(R-R_\odot)}{2.4\ {\rm kpc}}} e^{\frac{-|z|}{(509-48R({\rm kpc})\ {\rm pc})}} \frac{285\ {\rm pc}}{h_z(R)}
\label{flare}
,\end{equation}
a flared disc with a fixed scalelength of 2.4 kpc and a scaleheight that varies according to eq. (\ref{hz}).

Due either to a central hole in the stellar distribution or to a
flared disc that redistributes the stars in higher heights above the plane,  there is a
quite significant deficit of K2\,III stars 
in Galactic plane in the inner disc.
Therefore, although we must treat these results
with some care, we think that the present conclusions of this section
are to be considered as a tentative detection of a flat (rather than exponential) density 
distribution in the
inner disc of our Galaxy for near plane regions.
A remarkable aspect of this deficit of stars is that it affects mainly low $z$,
i.e.\ low-latitude regions, because the larger scaleheight
in the central regions compensates for the deficit of star for higher latitudes
(see Fig. \ref{Fig:densz}). 
This agrees with the fact that the deficit 
was not detected in Baade's windows (at $b=-3.9^\circ $) (Kiraga et al. 1997). 
High $z$ regions in the model do not have  a large deficit of K2\,III stars but 
even an excess with respect to the extrapolation of L02 outer
disc model (see Fig. \ref{Fig:densz} for $z=400$ pc).
Presumably, the deficit of stars affects
only the near-plane regions, where the disc material can
be swept up due to the motion of the in-plane bar (Athanassoula 1992, L01).

\section{Star counts in the near-infrared}

A flat density distribution 
in the inner disc would also produce a deficit of star counts
relative to an exponential disc.
Counts up to magnitude 9 in the $K$ band are representative of the old
stellar population and are appropriate for studying the central regions of the
Galaxy. Much deeper counts would make the local disc predominant instead
(Garz\'on et al. 1993).

\begin{figure}
\begin{center}
\mbox{\epsfig{file=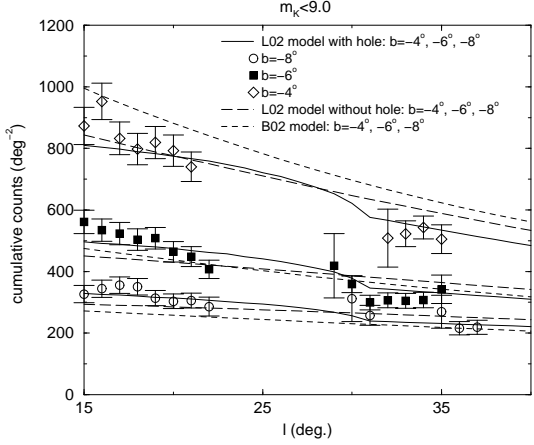,height=6cm}}
\end{center}
\caption{2MASS star counts with $m_{K}\le 9.0$ for $15^\circ <l<40^\circ $.
Comparison with B02 disc model, and the extrapolation of the L02 model of 
outer disc towards the centre with exponential and flat density distribution
(this paper) are shown. 
All models were normalized to give the best fit to the counts.}
\label{Fig:cuentasinner9}
\end{figure}

\begin{figure}
\begin{center}
\mbox{\epsfig{file=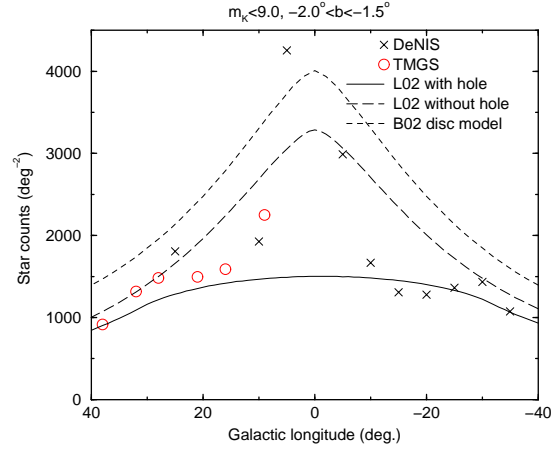,height=6cm}}
\end{center}
\caption{DENIS/TMGS star counts with 
$m_{K}\le 9.0$ for $-2^\circ <b<-1.5^\circ $ (L01). 
Comparison with  the B02 disc model, and the extrapolation of the L02 
model of outer disc towards the centre with an exponential and flat density 
distribution (this paper) are shown, with 
the same normalization as in  Fig.\ \protect{\ref{Fig:cuentasinner9}}.
Note that the models represent only the disc
(without the bulge, bar and ring) while the counts do include everything
in the line of sight.}
\label{Fig:DENIScounts9e}
\end{figure}

In Figs \ref{Fig:cuentasinner9} and \ref{Fig:DENIScounts9e} we
see how the model with a flat density distribution 
fits  the counts in off-plane regions quite well,
while the extrapolation of the L02 external disc model or the B02 disc model
(the luminosity function and the extinction were taken from the SKY model;
updated version of Wainscoat et al.\ 1992)
is significantly worse, especially for near-plane regions. 
All models were normalized to give the best fit to the counts presented
in Fig. \ref{Fig:cuentasinner9}. 
The extinction is very small, so it does not matter whether the applied
corrections are totally accurate.
No attempt is made to fit the plane ($z=0$) regions, where the in-plane
bar (L01) and the higher extinction make the conclusions 
about the disc less clear.

Since the L02 model is based in star counts, it represents
better than B02 (based on flux maps) the variation of counts with
latitude. L02 model also includes the
variation of $h_z(R)$ (a flare), making $h_z$ lower towards the
centre, improving the fit.
However, the L02 extrapolation is insufficient to fit the counts, especially
in $b=-1.75^\circ $ and also quite remarkably in $b=-4^\circ$; for higher
latitudes there are not so many differences. 
From here it seems clear that both the flare and
the inner truncation are necessary to fit the counts.

Obviously, in Fig. \ref{Fig:DENIScounts9e} we see a high excess of counts
in $|l|<10^\circ$ which is not reproduced by the disc model because it
is caused by the bulge. We see also some excess in positive longitudes with respect
to negative longitudes, probably due again to the in-plane bar (L01).
The most remarkable fact about the disc in Fig. \ref{Fig:DENIScounts9e}
is, as was said in L01,   that the
counts are constant between $|l|=15^\circ$ and $|l|=30^\circ$ 
(it is clearer at negative longitudes), and this cannot be
fitted with a purely exponential disc in the inner stellar disc.
COBE/DIRBE data at 2.2 $\mu $m also give a nearly flat light distribution, which 
at least does not decrease towards the centre, 
for $15^\circ <|l|<30^\circ $, $b=0^\circ$ (Hammersley et al.\ 1994, Fig. 1);
although in $b=0^\circ$ the imprint of other features such as the
in-plane bar or spiral arms also appears.

\subsection{Discussion of extinction effects}

For the near infrared counts, particularly for $|b|=1.75^\circ $,
where the deficit of counts in the central part is more evident
(see Fig.\ \ref{Fig:DENIScounts9e}), the extinction is not very high,
making it rather unlikely that this is a problem of the calculation
of the extinction.
For instance, at $|l|=15^\circ $ there is a deficit of 30--45\% in the
counts with respect to the L02 model without a flat inner density 
(and even more for the B02 model), which would need an error 
of 0.3--0.4 mag in extinction
in order to be explained in terms of dust.
With the model that we have used, the total extinction in the
Galactic inner disc at $l=15^\circ $, $b=1.75^\circ $
is 0.48 mag (up to a distance from the sun of 7.6 kpc). 
This is consistent with the estimate of the reddening 
for the red clump stars in the colour--magnitude diagrams (see \S \ref{.CMD})
at $l=15^\circ $, $b=2^\circ $, 
which gives $\approx$0.45 mag of total extinction up to that distance
(for $b=1.75^\circ $ it would be $\sim$0.06 mag larger).
Another independent estimate based on Schultheis et al.\ (1999)
gives $A_V\approx 5-6$ mag for regions in $b=-1.75^\circ$, 
$7.5^\circ <l<17.5^\circ$ (see Fig.\ 8 in L01),
i.e.\ 0.5--0.6 mag in K, smaller than the $\sim$0.8--0.9 mag or even more, 
necessary to produce the observed deficit of stars. 
A further difficult point, in case we suspected that the extinction
is responsible for this distribution of counts,
would be to explain the constant star counts 
in $15^\circ <|l|<30^\circ $. It is difficult to 
explain an excess of extinction precisely in a region where we know 
that there is a hole of gas/dust along the line of sight.
The possible extra extinction produced by a ring of gas of radius
$\sim$3--4 kpc would give
less extinction for lower $|l|$, because the crossing of the line of sight
in the ring is more perpendicular to the ring than tangential; 
and, moreover, the ring is likely to have a significant 
contribution only for $|b|<1^\circ $ (L01).
Therefore,  
we feel that many coincidences that  would be difficult to explain would need
to  take place in order to explain
the observed facts in terms of  only a problem of extinction calculation.

\begin{figure}
\begin{center}
\mbox{\epsfig{file=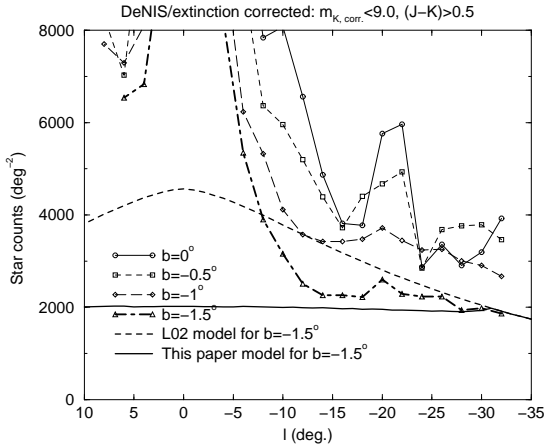,height=6cm}}
\end{center}
\caption{DENIS star counts with $m_{K,{\rm corrected}}\le 9.0$ and $(J-K)>0.5$
based on L01-Fig. 10 (see method of extinction correction in L01, 
using $(J-K)$ colors). $\Delta l=2^\circ $, $\Delta b=0.5^\circ $.}
\label{Fig:plotestat6b_2D_b0-1_5}
\end{figure}

Examination of the counts corrected
for extinction is a further test that can be carried out. 
In Fig.\ 10 of L01, we can see the counts for the negative
latitudes after they were approximately corrected for extinction (see L01 for
details). The colour $(J-K)>0.5$ was chosen  to remove part of
the local disc sources and improve the contrast of distant sources.
In Fig. \ref{Fig:plotestat6b_2D_b0-1_5} we reproduce the
counts of L01 (fig.\ 10) as a function of longitude for 
four different latitudes.
In this figure, 
we  observe first of all that the stellar 3 kpc arm (that is, the ring,
at $l\approx -22^\circ $) 
affects only those counts for $|b|<1^\circ$. 
Second, and more importantly,
we can see that the counts are nearly longitude--independent 
for the range $-32^\circ <l<-16^\circ $ at $b=-1.5^\circ $. This is also true
for latitudes closer  to the plane, but the 3 kpc arm at $l\approx -22^\circ $
is added. Therefore, we see again that the shape of the counts is 
incompatible with an exponential disc without truncation, which would
give increasing counts towards the centre; in this case, we 
are talking about counts corrected for extinction.
Again, many unlikely coincidences would need to occur
 to explain the observed facts without the disc truncation.

\section{Density distribution of OH/IR stars at 1612 MHz through  
kinematical distance determination}
\label{.OHIR}

Stars with high mass-loss rates produce OH masers whose strongest
line is at 1612.23 MHz (Elitzur et al.\ 1976). This allows us to observe
these stars at this radio frequency with the advantage of 
negligible extinction. A complete survey of OH/IR stars
in the region $|l|<45^\circ $, $|b|<3.25^\circ $, 
  ATCA/VLA
OH 1612 MHz, was done by 
Sevenster et al.\ (1997a, 1997b, 2001). These sources trace the intermediate--old population 
when the sources with wind outflow velocity ($v_e$) less than 14 km s$^{-1}$
are selected, as will be
done in this paper; they have an age of $\approx 1.5-8$ Gyr (Sevenster 1999).
The completeness is very close to 100\% for
sources with flux larger than 0.375 Jy (Sevenster et al. 1997a); 
therefore, we use only sources above this limit in this paper.

Star counts cannot be used as a test to discriminate 
among the models since the number of sources is very small and
the Poisson errors dominate.
Another direct way to check whether there is a deficit of sources at
$R<4$ kpc is to obtain the distances
of the sources. The survey has information on the radial velocities
of the sources so, in principle, it is possible to carry out
such a test. The only difficulty is that the number of sources is
not very high, although high enough to discriminate among the different
models. 
The test is performed as follows: first we calculate the galactocentric
distance of each source in certain regions of the sky (which contain
inner disc sources but avoiding regions where other components
contribute a large number of sources); 
second, we count the number of sources
in the inner disc region in successive radial rings; 
third, we compare the results with the predictions of different density models
using a luminosity function that we also obtain directly from
the data.

\subsection{Galactocentric distance determination}

The first step, the galactocentric distance ($R$)
determination, is carried out through the well known relation between
radial velocity, $v_r$, and $R$ for in-plane regions (Burton 1988, eq.\ 7.4),
assuming circular orbits:

\begin{equation}
v_r=R_\odot \left[\frac{V(R)}{R}-\frac{V(R_\odot)}{R_\odot}\right]
\sin l
,\end{equation}
where $V(R)$ is the rotation velocity of the disc at galactocentric
distance $R$.
If we take $V(R)=V(R_\odot)(R/R_\odot )^{0.1}$ km s$^{-1}$ (Binney et al. 1991), 
this leads to

\begin{equation}
R=\frac{R_\odot}{\left(1+\frac{v_r}{V(R_\odot)\sin l}\right)^{1.11}}
\label{Rkine}
.\end{equation}
This equation is applicable because the disc OH/IR stars in the plane
follow the galactic rotation (Baud et al.\ 1981). 
We adopt $V(R_\odot)=200$ km s$^{-1}$ (Honma \& Sofue
1996) (a value of 220 km s$^{-1}$ would not change the conclusions of this
section, but would slightly
reduce the values in Fig.\ \ref{Fig:ratioOHIR}/Table \ref{Tab:ratioOHIR}) 
and $R_\odot=7.9$ kpc (L\'opez-Corredoira et al.\ 2000). 
The random motions in the old population 
are around 30--40 km s$^{-1}$ (Baud et al.\ 1981). 
$\sigma_{v_r}\approx 35$ km s$^{-1}$ leads to an error of 

\begin{equation}
\sigma _R\approx \frac{0.020}{\sin l}R({\rm kpc})^{1.9}\ {\rm kpc}
\label{erR}
,\end{equation} 
and this leads to important errors only for large values of $R$ 
and low $l$. Moreover, the errors for the
individual sources are mostly reduced when the statistical count
is carried out, except perhaps for the very few sources close to
the Sun (high $R$). In any case, we will include the calculation
of this error in this paper  to show that even with this
error (and the Poisson errors) models with a purely 
exponential disc are excluded.

\subsection{Luminosity function}

The previous relation allows us to determine $R$, but not the distance from the
Sun ($r$) because of the well-known ambiguity of there being two possible values for a given
$R$ inside the solar circle (Burton 1988). 
Therefore, we cannot use this relation to obtain 
the luminosity function directly. 

\begin{figure}
\begin{center}
\mbox{\epsfig{file=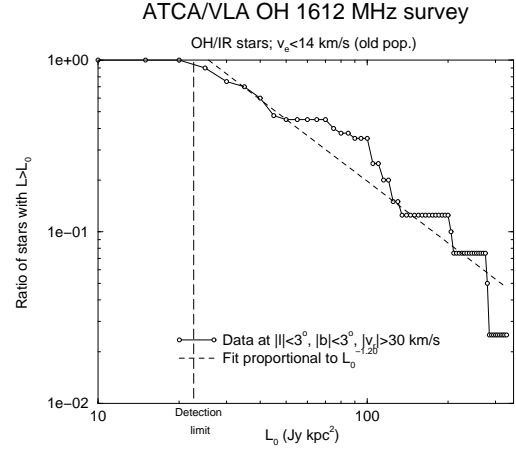,height=6cm}}
\end{center}
\caption{Luminosity distribution of ATCA/VLA-1612 MHz survey sources in the
centre of the Galaxy ($|l|<3^\circ $,$|b|<3^\circ $) and with radial
velocities $|v_r|>30$ km s$^{-1}$.}
\label{Fig:lumOHIR}
\end{figure}

To estimate the luminosity function, we take separately the sources in
the centre of the Galaxy ($|l|<3^\circ $, $|b|<3^\circ $) and assume
that all of them are at the same distance ($R_0=7.9$ kpc).
The contribution of foreground disc sources is small.
To reduce this disc contamination further, we take the sources with radial
velocities $|v_r|>30$ km s$^{-1}$,
since the disc sources in this direction are expected to have near-zero 
radial velocities.
Fig.\ \ref{Fig:lumOHIR} shows the cumulative luminosity distribution 
of this subsample (luminosity
in units Jy kpc$^2$, i.e. the flux of the source in Jy considering the source
located at a distance of 1 kpc) which can be fitted by:

\begin{equation}
\int _{L_0}^\infty \phi(L)dL\propto L_0^{-1.20}
,\end{equation}
that is, a luminosity function $\phi (L)\propto L^{-2.20}$.
This relation is fitted for $L_0>0.375R_\odot ({\rm kpc})^2$ 
Jy kpc$^2$, the detection limit. 
The exponent, $\alpha =-2.2$, is close to the value of $-$2.0 obtained
also in the Galactic center by Herman \& Habing (1985).
Chengalur et al.\ (1993) obtained a value of $-$1.5 and Baud et al.\ (1981)
obtained a value of $-$1.65, but their samples also contained
young population (high $v_e$), so the difference could stem from this.
We are interested here only in the old population (older than 1.5 Gyr).
A power law is generally accepted as a fit to the luminosity function, at least in the observed range. 

There is a cut-off for the faintest luminosities (Baud et al.\ 1981), which
we take to be at $L_0<0.2R_\odot({\rm kpc})^2$ Jy kpc$^2$, 
since the luminosity function peaks around $L_0=0.2R_\odot ({\rm kpc})^2$ 
Jy kpc$^2$ (Sjouwerman et al.\ 1998), and the number of fainter sources 
is lower (so the factor $r^2$ will make 
the number of nearby faint sources negligible).
Debattista et al.\ (2002) have also verified that the number of low
luminosity sources is very small 
(corresponding to low $v_e$; luminosity is correlated
with $v_e$ but the relation between both parameters cannot be used
as a distance indicator since the dispersion of luminosities for a given
$v_e$ is large).
Fig.\ 9 in Debattista et al.\ (2002) shows that there are
practically no low-luminosity stars ($v_e<6$ km s$^{-1}$; roughly less than
$0.2R_\odot({\rm kpc})^2$ Jy kpc$^2$ in average) 
at $20^\circ<|l|<45^\circ$ in the ATCA/VLA survey.
Also, Sevenster et al.\ (1997a, fig.\ 6b; 2001, fig.\ 5b) find a very
low number of low-$v_e$ sources in the whole survey. 

\begin{figure}
\begin{center}
\mbox{\epsfig{file=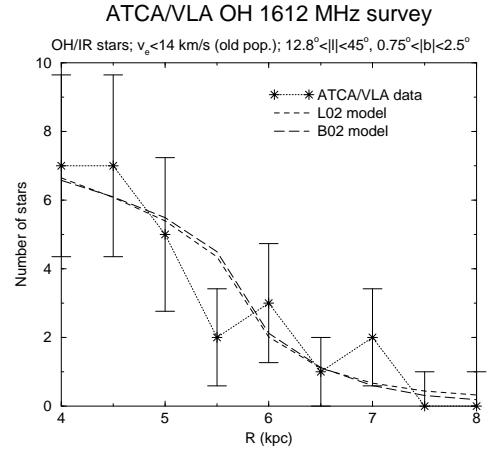,height=6cm}}
\end{center}
\caption{Distribution of ATCA/VLA-1612 MHz survey sources (flux larger 
than 0.375 Jy, outflow velocities $v_e<14$ km s$^{-1}$) in the
region $12.8^\circ<|l|<45^\circ $, $0.75^\circ<|b|<2.5^\circ$ as a
function of the galactocentric radii for $R>4$ kpc. 
Model normalized to give the same number of sources than the data.
Error bars stand only for Poisson errors.
There is a good agreement with the B02 or L02 models (it does not matter whether
it has a flat distribution or not, since this does not affect 
the regions at $R>4$ kpc) in this range of $R$.}
\label{Fig:OHIR_R}
\end{figure}

With the luminosity function and the densities of the L02 or B02 models,
we can calculate the number of counts in a given region
of the sky as a function of $R$, and compare them with the
distribution obtained from the data through eq. (\ref{Rkine}).
Fig. \ref{Fig:OHIR_R} shows the result for $R>4$ kpc in 
$12.8^\circ<|l|<45^\circ $. This figure shows an
agreement between models and data at $R>4$ kpc which corroborates that the
luminosity function is approximately correct.

\subsection{Deficit of stars in the inner disc}

\begin{figure}
\begin{center}
\mbox{\epsfig{file=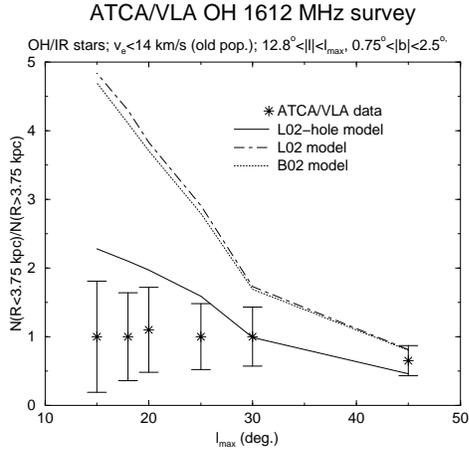,height=6cm}}
\end{center}
\caption{Distribution of ATCA/VLA-1612 MHz survey sources (flux greater 
than 0.375 Jy, outflow velocities $v_e<14$ km s$^{-1}$) in the
region $12.8^\circ<|l|<l_{\rm max}$, $0.75^\circ<|b|<2.5^\circ$. 
Note that the points are not independent, however the 
point at $l_{\rm max}=18^\circ $ is enough to reject the B02 
and L02 models beyond the 5$\sigma $ level.}
\label{Fig:ratioOHIR}
\end{figure}

If we apply this technique to the inner disc regions, 
we can see whether the model of \S \ref{.CMD} does or does not fit
better than the L02 or B02 exponential models.
We select the region $|l|>12.8^\circ$ (i.e. $R>1.75$ kpc), 
$0.75^\circ<|b|<2.5^\circ$
because this is far enough away from the bulge; it crosses the inner disc
and is far enough away from the plane to
consider the contribution of other possible plane old-population
components (bar, ring) small in comparison to the error bars
of the data. Rather than a plot as a function of $R$ (as in Fig.\
\ref{Fig:OHIR_R}), 
we will show the ratio between lower and higher $R$ sources, between
the sources in the inner disc ($R<3.75$ kpc) and the remainder.
The Poisson errors are very large if we divide the data
into bins with different $R$. 
By means of eq. (\ref{Rkine}),
the ratio $\left(\frac{N(R<3.75\ {\rm kpc})}{N(R>3.75\ {\rm kpc})}\right)
=\left(\frac{N([v_r/\sin l]>190\ {\rm km/s})}
{N([v_r/\sin l]<190\ {\rm km/s})}\right)$ is
shown in Table \ref{Tab:ratioOHIR}.
The lower the maximum longitude, the more sensitive we are
to the inner disc population; however, the Poisson
errors are larger (because we have fewer sources).
Restricted to $|l|<18^\circ$, we get the larger discrepancy:
the differences of an exponential model and the results from the data are 
around 5$\sigma $, significant enough. 
Figure \ref{Fig:ratioOHIR} shows this graphically.

The model of this paper gives a much better fit:
all the ranges in Table \ref{Tab:ratioOHIR} are within the 1.8$\sigma$ level.
Perhaps the deficit of stars should be larger than that calculated
in \S \ref{.CMD}, but since we have not taken into account the errors
of eq. (\ref{hole}) and the interpolation of the transition
region, the compatibility with the flat distribution described in this paper 
is not excluded.

\begin{table*}
\caption{The $\left(\frac{N(R<3.75\ {\rm kpc})}{N(R>3.75\ {\rm kpc})}\right)$ ratio
 in the ATCA/VLA 1612 MHz survey
and the predictions of three different models. Range of
latitudes: $0.75^\circ <|b|<2.5^\circ$. Errors in the data include
Poisson errors and the errors due to the dispersion of values
in $R$ calculated  according to eq.\ (\protect{\ref{erR}}) [estimated
by means of a Monte Carlo simulation].}
{\centering \begin{tabular}{ccccc}\hline 
\hline
Long. range & Data & Model ``This paper''  & Model L02 & Model B02\\ \hline 
$12.8^\circ<|l|<45^\circ$ & 0.65$\pm $0.22 & 0.46 & 0.81 & 0.80 \\
$12.8^\circ<|l|<30^\circ$ & 1.00$\pm $0.43 & 0.99 & 1.73 & 1.69 \\
$12.8^\circ<|l|<25^\circ$ & 1.00$\pm $0.48 & 1.59 & 2.91 & 2.79 \\ 
$12.8^\circ<|l|<20^\circ$ & 1.10$\pm $0.62 & 1.97 & 3.83 & 3.71 \\
$12.8^\circ<|l|<18^\circ$ & 1.00$\pm $0.64 & 2.10 & 4.30 & 4.11 \\
$12.8^\circ<|l|<15^\circ$ & 1.00$\pm $0.81 & 2.28 & 4.84 & 4.70 \\ \hline
\label{Tab:ratioOHIR}
\end{tabular}\par}
\end{table*}

Therefore, the conclusion of this section is again that a deficit of stars
is necessary in the old population (older than 1.5 Gyr) of the
disc with a density amplitude similar to that derived
in \S \ref{.CMD}, or even lower.
It is noteworthy that this deficit of old population
OH/IR stars has already been argued by Baud et al.\ (1981) for the populations
around 1 Gyr old.
We quote a paragraph from \S 6.4 of this paper\footnote{Baud et al.\ call  
the separation between the two peaks of OH/IR sources at 1612 MHz
$\Delta V=2v_e$ and use $R_\odot=10$ kpc.}:
\begin{quote}
That the sources with large $\Delta V$ exhibit the same 
kind of overall density distribution is to be expected when they are
as young as we have estimated. But the similarity of the density
distribution of the sources with small $\Delta V$($\approx 10^9$ yr
old) and the extreme Population I objects is surprising. We interpret
this to mean that the large-scale CO distribution in the disc and
hence the region of active star formation has not changed significantly
during the last $10^9$ yr, or 4-8 rotations of the Galaxy. Therefore the
`hole' in the gas and in the OH/IR sources distribution at $R<4.5$ kpc and
the maximum density at $R=5$ kpc appear to be rather stable phenomena.
\end{quote}
We corroborate these words for even
older stars.

\section{Counts with bright sources in the mid-infrared}

In this section, we compare the disc models
with bright mid-infrared star counts in the plane (i.e.\ a young population).
In the mid-infrared, there is practically zero extinction 
($\sim$70 smaller than in $V$) even in plane 
regions. This means
that  extinction in the plane (b=0$^\circ$) is typically 0.1--0.2 mag in 
the mid-infrared, except in
the very centre, where it can reach 0.4 mag.

\begin{figure}
\begin{center}
\mbox{\epsfig{file=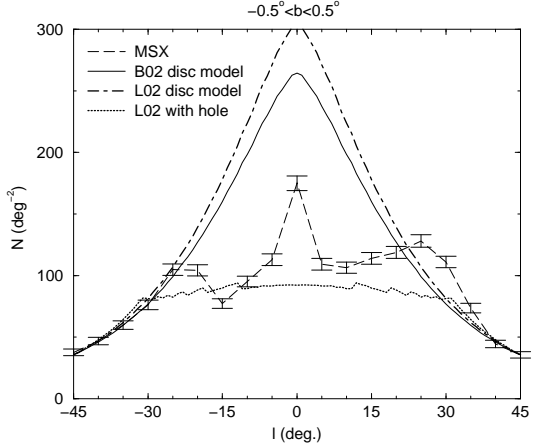,height=6cm}}
\end{center}
\caption{MSX star counts in the plane at 14.6 $\mu$m
up to magnitude 3.0. Comparison with the B02 
disc model and the extrapolation of the L02 model of outer disc 
towards the centre with an exponential and flat density distribution
(this paper). All models were normalized
with the counts to fit the counts at $l=45^\circ$.
It is clear that the deficit of stars in the young stellar population represented
here is necessary.
Note that the models represent only the disc
(without a bulge, bar or ring) while the counts include everything
in the line of sight.}
\label{Fig:MSX0}
\end{figure}

The comparison is carried out
in Fig.\ \ref{Fig:MSX0}. In this case, MSX counts at 14.6 $\mu$m
up to magnitude 3.0 are compared with different models.
This is compared with the B02 model prediction
(the luminosity function and the extinction were taken from the SKY model;
updated version of Wainscoat et al.\ 1992) and with the extrapolation
of the outer disc according to L02. 
Obviously, the L02 and B02 models fail to reproduce the data for $|l|<30^\circ$.
COBE/DIRBE data at 12 $\mu $m, representative of the flux of
the young population, also give a nearly flat light distribution 
at $10^\circ<|l|<30^\circ $, $b=0^\circ$ (Hammersley et al.\ 1994, fig. 1) instead of 
increasing towards $l=0^\circ$.
Comparison with the models shows that the inner truncation of
the exponential increase in the density in the disc
can explain the difference. The model of this paper, which follows
eq.\ (\ref{hole}) for $R<4$ kpc, can 
roughly fit the counts. 
Errors in the extinction are very low and cannot 
explain the observed difference between data and exponential models. 
The bulge, spiral arms, ring and in-plane bar also have 
some contribution which is not calculated here; note, however, that they would
increase the number of counts but never decrease it, so the interpretation of
a deficit of stars is unavoidable. 
No attempt is made to fit the counts for
other latitudes because we would need the scaleheight and the
flare of the observed population with MSX-14.65 $\mu$m, dominated
by a young population, and this is at present unknown.

\section{Summary of results: a smooth model for outer and inner disc}

\begin{figure}
\begin{center}
\mbox{\epsfig{file=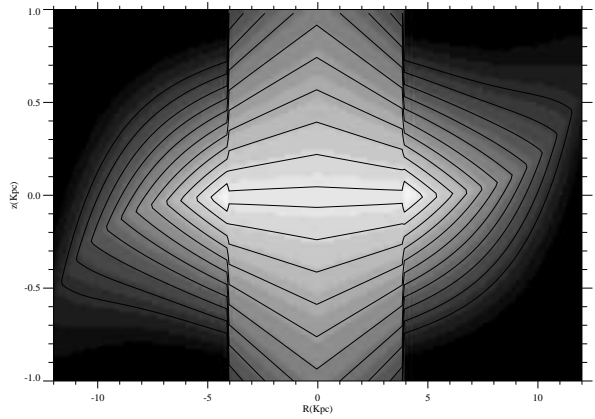,height=8cm,angle=90}}
\end{center}
\caption{Contour diagram of $\log _{10}\rho $ (pc$^{-3}$) 
in the $yz$-plane of the Galaxy (perpendicular to 
the line Sun--Galactic centre) with $y$ between $-$12.0 and 12.0 
kpc and $z$ between $-$1.0 and 1.0 kpc (vertical scale in the plot 
multiplied by a factor 5). Lower contour: 
$\log _{10}\rho $ (pc$^{-3}$)=-2.1; step=0.15.
Model: L02 for $R>4$ kpc, and eq. (\ref{hole}) for 
$R\le 4$ kpc. The abrupt transition (at $|y|=R=4$ kpc) is due to the
discontinuous change of regime between the inner disc and the outer disc.
Note also that the model of this paper is valid only for regions not
very far from the plane.} 
\label{Fig:contourinner}
\end{figure} 

\begin{figure}
\begin{center}
\mbox{\epsfig{file=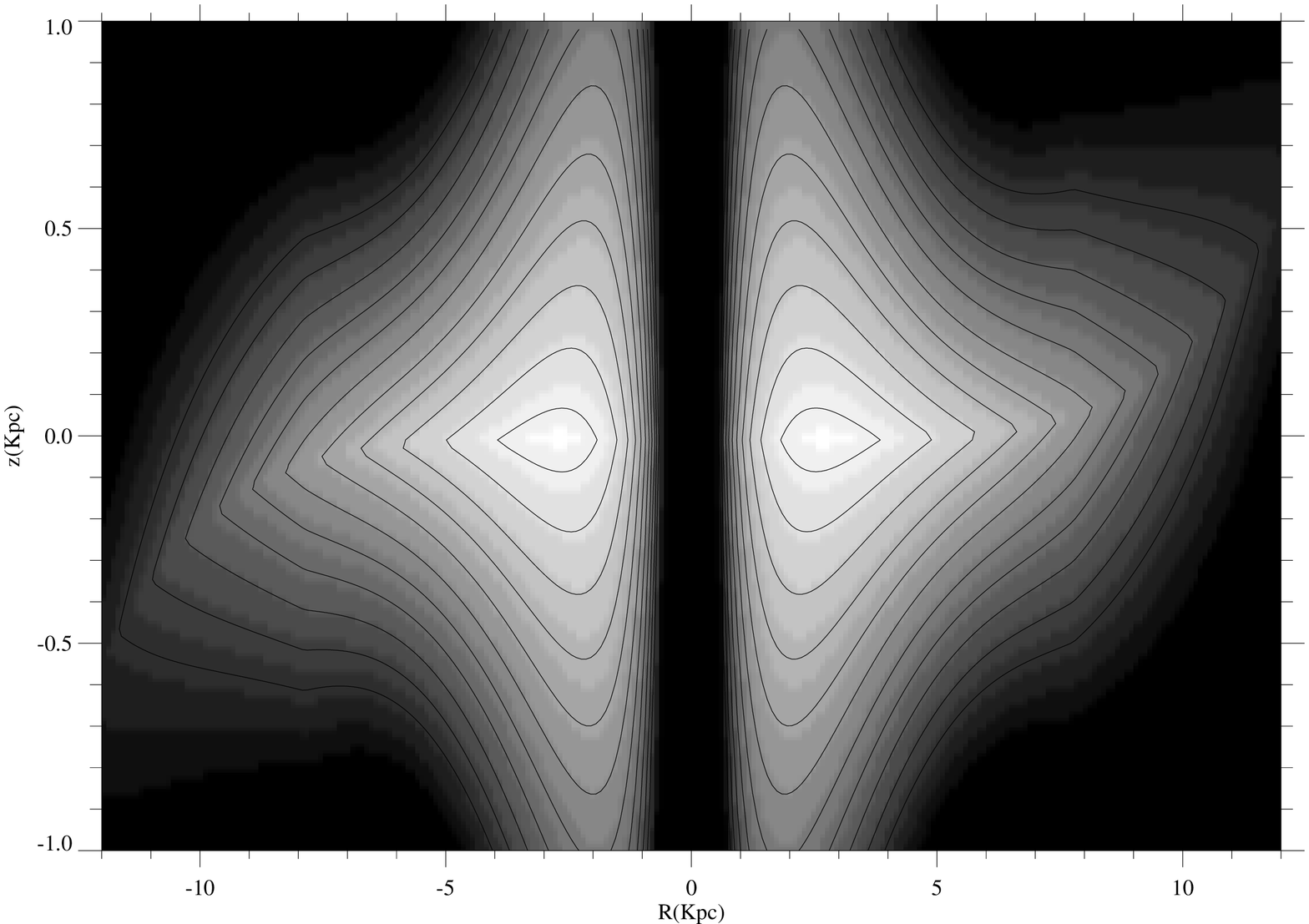,height=8cm,angle=90}}
\end{center}
\caption{Contour diagram of $\log _{10}\rho $ (pc$^{-3}$) of a possible
interpolated/extrapolated smooth model of the disc according to
eq. (\ref{expdob}) for $R<R_\odot $ and L02 for $R\ge R_\odot $
in the $yz$-plane of the Galaxy (perpendicular to 
the line Sun--Galactic Centre) with $y$ between $-$12.0 and 12.0 
kpc and $z$ between $-$1.0 and 1.0 kpc (vertical scale in the plot 
multiplied by a factor 5). Lower contour: 
$\log _{10}\rho $ (pc$^{-3}$) = $-$2.1; step = 0.15.} 
\label{Fig:contourinner2}
\end{figure}

We have used three different methods: 1) the inversion of the red clump giant 
distribution in near-infrared colour--magnitude diagrams  to obtain 
the star density along the line of sight; 2) the determination of the density 
distribution of 1612 MHz sources by means of the distance 
determination of OH/IR sources from their 
kinematical information; 3) the analysis of near-infrared star counts and
comparison with models. All the tests agree with the following picture
for the inner disc:

\begin{itemize}

\item There is a deficit of stars with respect to an exponential disc 
in the Milky Way's inner stellar disc for 2.25 kpc $<R<4$ kpc.
The density in the plane ($b\approx 0^\circ$, i.e. $z\approx 0$)
is almost independent of $R$ instead of being an exponential law 
of the type $\rho \propto \exp (-R/h)$.

\item There is a slight increase in scaleheight in the
central disc, $h_z=509-48R({\rm kpc})$ pc, 
so the combined effect of deficit of stars in the plane 
and higher scaleheight produces an important deficit 
of stars for low latitude regions, although not for high latitude regions.
Roughly, the density follows an expression of the type 
$\rho (R,z)\approx 5.1\pm 0.6\rho_\odot e^\frac{-|z|}{509\pm 67-48\pm 20
R(kpc)\rm \ pc}$, a result that is also compatible with a flared inner disc with
a fixed scalelength of 2.4 kpc: $\rho (R,z)\approx\rho_\odot e^\frac{-(R-R_\odot)}{2.4 {\rm
kpc}} e^\frac{-|z|}{509\pm 67-48\pm 20
R(kpc)\rm \ pc} \frac{285\ {\rm pc}}{509\pm 67-48\pm 20
R(kpc)\rm \ pc}$. It is our objective to investigate further the real nature of the inner
stellar disc, with a more complete analysis of 2MASS \emph{All Sky Release} star counts. 
See the graphical representation in Fig.\ \ref{Fig:contourinner} (both possibilities
 are numerically equivalent).
Note that these numbers are also subject to several systematic 
errors ($\la$ 15\% in the density between latitudes 
2 and 6 degrees), so the parameters might change; although not 
as much as to allow the fit of a purely exponential disc 
(which would require errors greater than 70\%). Moreover, the deficit of
stars is corroborated with the OH/IR sources and the star counts.
These parameters are applicable only for regions
not very far from the plane ($|b|<6^\circ $).

\item Both the 
young and old population have this deficit of stars, and at present
no significant differences in their distributions have 
been deduced from the data. 
Presumably, a more accurate measure of the distribution might reveal
some differences. At present we can say
that, roughly speaking, the deficit of stars with respect the extrapolation of
an exponential distribution
is probably a rather stable feature of the disc, 
which might be due to the existence of an in-plane bar that sweeps out the 
near-plane stars.

\item An expression that summarizes the L02 outer disc ($R>6$ kpc)
and the inner disc (2.5 kpc $<R<4$ kpc) with a 
smooth transition between two regimes is:

\[
\rho(R,z)\approx \left[\rho _\odot e^{\left(\frac{R_\odot}{1970\ pc}+
\frac{3740\ pc}{R_\odot}\right)}\right]  
\]\begin{equation}
\times e^{-\left(\frac{R}{1970\ pc}+\frac{3740\ pc}{R}\right)}e^{-|z|/h_z},
\label{expdob}
\end{equation}

with
\[
h_z(R)\approx285[1+0.21\ {\rm kpc}^{-1}(R-R_\odot)
\]\begin{equation}
+0.056\ {\rm kpc}^{-2}(R-R_\odot)^2] \ {\rm pc}
.\end{equation}

This density is plotted in Fig. \ref{Fig:contourinner2}. 
Although this expression
was not tested for all ranges (for instance, it is not tested around $R=5$ kpc
or for $R<2$ kpc), this can be used as an approximate 
interpolation/extrapolation that models the disc in $R<R_\odot $.
For the outermost disc, it is better still  to use the expression
of L02. The innermost disc ($R<2$ kpc) reveals with the 
expression (\ref{expdob}) a very deep hole. Unfortunately, we cannot
be sure of the existence of
this deep hole there, since the disc and the bulge cannot
be easily separated in that region. It is clear at least that
between 2 and 4 kpc the deficit of stars is very significant.

\end{itemize}

\

Acknowledgments:
We thank C. Alard, V. P. Debattista and P. L. Hammersley.
This publication makes use of data products from 2MASS (which is
a joint project of the Univ. of Massachusetts and the Infrared Processing
and Analysis Center (IPAC), funded by the NASA and the NSF);
the MSX Point Source Catalog (which was obtained from the NASA/IPAC Infrared 
Science Archive at Pasadena, California); DENIS (the
result of a joint effort involving personnel and financial
contributions from several Institutes, mostly located in Europe
and supported financially mainly by the French Institute National 
des Sciences de l'Univers, CNRS, and French Education Ministry, 
the European Southern Observatory, the State of Baden-W\"urttemberg, 
and the European Commission under a network of the Human Capital and Mobility 
programme), the TMGS (based on observations made at Carlos S\'anchez
Telescope operated
by IAC at Teide Observatory on the island of Tenerife),
the  ATCA/VLA OH 1612-MHz survey (based on observations made by Sevenster
et al.\ at the Australia Telescope Compact Array and at the Very Large Array).

\end{document}